\documentclass[aps,prb,amsmath,amssymb,nobibnotes]{revtex4}

\usepackage{bm}
\usepackage{latexsym}
\usepackage{amsthm}
\usepackage{tensor}

\newtheoremstyle{jmp}
{\topsep}{\topsep}
{\normalfont\itshape}
{\parindent}{\normalfont\bfseries}
{:}{.5em}
{}

\theoremstyle{jmp}
\newtheorem{dfn}{Definition}
\newtheorem{prop}{Proposition}

\newcommand{\vect}[1]{\bm{#1}}

\makeatletter
\renewcommand\frontmatter@title@above{}
\renewcommand\frontmatter@title@format{\large\bfseries\raggedright}
\renewcommand\frontmatter@title@below{\addvspace{6\p@}}
\makeatother

\makeatletter
\renewcommand\frontmatter@authorformat
{\preprintsty@sw{\vskip0.5pc\relax}{}
\@tempskipa\@flushglue
\@flushglue\z@ plus50\p@\relax
\raggedright\advance\leftskip20mm\relax
\@flushglue\@tempskipa
\parskip\z@skip}
\makeatother

\makeatletter
\renewcommand\frontmatter@affiliationfont
{\small\slshape\selectfont\baselineskip10.5\p@\relax
\@tempskipa\@flushglue
\@flushglue\z@ plus50\p@\relax
\raggedright\advance\leftskip20mm\relax
\@flushglue\@tempskipa}
\makeatother

\makeatletter
\renewcommand{\section}{\@startsection{section}{1}{\z@}
{0.8cm \@plus1ex minus .2ex}{0.5cm}{\normalfont\small\bfseries}}
\renewcommand{\subsection}{\@startsection{subsection}{2}{-12pt}
{0.8cm \@plus1ex minus .2ex}{0.5cm}{\normalfont\small\bfseries}}
\renewcommand{\subsubsection}{\@startsection{subsubsection}{3}{-12pt}
{0.8cm \@plus1ex minus .2ex}{0.5cm}{\normalfont\small\itshape}}
\makeatother

\makeatletter
\renewcommand\bibsection
{\section*{}
\@nobreaktrue}
\makeatother

\begin{document}

\title{Trkalian fields and Radon transformation}

\author{K. Saygili}
\email{ksaygili@yeditepe.edu.tr}
\affiliation{Department of Mathematics, Yeditepe University, 
             Kayisdagi, 34755 Istanbul, Turkey}

\begin{abstract}
\noindent We write the spherical curl transformation for Trkalian fields 
using differential forms. Then we consider Radon transform of these fields. 
The Radon transform of a Trkalian field satisfies a corresponding eigenvalue 
equation on a sphere in transform space. The field can be reconstructed using 
knowledge of the Radon transform on a canonical hemisphere. We consider 
relation of the Radon transformation with Biot-Savart integral operator 
and discuss its transform introducing Radon-Biot-Savart operator. The Radon 
transform of a Trkalian field is an eigenvector of this operator. We also 
present an Ampere law type relation for these fields. We apply these to 
Lundquist solution. We present a Chandrasekhar-Kendall type solution of 
the corresponding equation in the transform space. Lastly, we focus on 
the Euclidean topologically massive Abelian gauge theory. The Radon 
transform of an anti-self-dual field is related by antipodal map on 
this sphere to the transform of the self-dual field obtained by inverting 
space coordinates. The Lundquist solution provides an example of quantization 
of topological mass in this context.

\end{abstract}

\maketitle

\setlength{\parindent}{4mm}

\section{INTRODUCTION}
\label{Intro}

  The eigenvectors of the curl operator are called Beltrami fields. 
These are specifically called Trkalian if the eigenvalue is constant. 
These arises in different areas ranging from fluid dynamics and plasma 
physics to field theories.

  The field theoretic examples of Trkalian vectors are the Euclidean 
topologically massive\cite{DJTS1, DJTS2, DJTS3, KS2, KS3, KS1} Abelian gauge 
fields and force-free\cite{GM} magnetic fields. The topologically massive 
gauge theories\cite{DJTS1, DJTS2, DJTS3} are qualitatively different from 
Yang-Mills type gauge theories besides their mathematical elegance and 
consistency. In this context, Trkalian type solutions on $3$-sphere 
$\mathbb{S}^{3}$, anti-de Sitter space $\mathbb{H}^{3}$ and other spaces 
in connection with contact geometry are discussed in Refs. 
\onlinecite{KS2, KS3, KS1}.

  The Curl transformation, as an interesting tool, is introduced in 
Ref. \onlinecite{M1}. The (spherical) curl transformation is developed 
for force-free magnetic fields using vectors.\cite{Ml1, Ml3} The spherical 
curl transform and Radon transform are equivalent descriptions.\cite{Ml3} 
More precisely, the spherical curl transformation is a Radon probe 
transformation.\cite{NW} Then the spherical curl transformation is 
applied to the Euclidean topologically massive Abelian gauge theory 
on $\mathbb{R}^{3}$ in Ref. \onlinecite{KS1}. A complex spinor 
formalism is also developed in Refs. \onlinecite{H, HL}.

  The Radon transformation and other variants of it have provided valuable 
insight into problems in different areas ranging from tomography to twistor 
theory. The Radon transformation can be defined for differential forms 
of certain orders in various dimensions.\cite{GGG, P1}

  In the next section, we shall first write the spherical curl 
transform for Trkalian fields using differential forms. Then we shall 
discuss Radon transform of these fields in its own right. The Radon 
transform of a Trkalian field satisfies a corresponding eigenvalue 
equation on a sphere in transform space whose radius is determined 
by the eigenvalue. This correspondence can also be inverted. The 
field can be reconstructed using knowledge of the Radon transform 
on a \textit{canonical} hemisphere.\cite{MP} 

  In Section \ref{RBS}, we shall consider the relation of the Radon 
transformation with Biot-Savart\cite{CDG} ($BS$) integral operator.
We shall also present an Ampere law type relation\cite{KS2, KS3} for 
the Trkalian fields. Then we shall discuss Radon transform of the $BS$ 
integral introducing Radon-Biot-Savart ($RBS$) operator in the transform 
space. The Radon transform of a Trkalian field is an eigenvector of the 
$RBS$ operator.

  In Section \ref{Sol}, we shall first present applications of our 
constructions on Lundquist\cite{L} (L) solution. Then we shall briefly 
present Chandrasekhar-Kendall\cite{CK} (CK) method with circular and 
elliptic cylindrical solutions. We shall make use of an analogous method 
for finding solutions of the corresponding equation in the transform space. 
This yields an eigenvector of the $RBS$ operator.

  The Trkalian fields yield interesting solutions of not only the 
topologically massive gauge theories\cite{KS2, KS3, KS1} but gravity\cite{ANS}
as well. We shall focus on the Euclidean topologically massive Abelian gauge 
theory in Section \ref{tmgf}. The Radon transform of an 
\textit{anti-self-dual} potential (or field) is related by antipodal 
map on the sphere to the transform of the \textit{self-dual} potential 
obtained by inverting space coordinates. 

  To the knowledge of the author, L and CK solutions have been overlooked 
in the context of topologically massive gauge theories. The L solution 
provides an example of quantization of the topological 
mass\cite{DJTS1, DJTS2, DJTS3, KS2, KS1} in this context. 

  We provide the necessary definitions and proofs in an appendix.

\section{The Radon transformation of Trkalian fields}

\subsection{The spherical curl transformation}
\label{Curltrf}

  We shall write the spherical curl transformation using differential forms 
in this subsection. The formulation in differential forms follows the same 
line of reasoning with formally analogous manipulations as given for 
vectors.\cite{M1, Ml1, Ml3}

\subsubsection{The Moses coframe and the eigenbasis}

  We use the usual correspondence between vectors and $1$-forms:
$\vect{Q}_{\lambda}(\vect{k}) \leftrightarrow 
\omega^{a}(\vect{k})$ on $\mathbb{R}^{3}$. We shall also make use 
of the correspondences $\vect{\nabla}\bm{\times}\leftrightarrow *d$, 
$\vect{\nabla}\cdot \leftrightarrow d*$, dot product: 
$(\cdot) \leftrightarrow (\wedge *)$, cross product: 
$(\bm{\times}) \leftrightarrow *(\wedge)$  between basic 
operations on vectors and differential forms. 
 
  The Moses\cite{M1} eigenbasis $1$-forms 
 
\begin{eqnarray}
\chi^{a}(\vect{x}|\vect{k}) 
=\frac{1}{(2\pi)^{3/2}} e^{i \vect{k} \cdot \vect{x}}
\omega^{a}(\vect{k}),
\end{eqnarray}

\noindent of operator $*d$ 

\begin{eqnarray}
& & *d \chi^{a}(\vect{x}|\vect{k})
=\lambda k \chi^{a}(\vect{x}|\vect{k}), 
\nonumber \\
& & d* \chi^{a}(\vect{x}|\vect{k})=0,
\hspace*{5mm} a=1, 2, \\
& & d* \chi^{3}(\vect{x}|\vect{k})
=-\frac{1}{(2\pi)^{3/2}} \,\, ik \, 
e^{i \vect{k} \cdot \vect{x}} *1,
\nonumber 
\end{eqnarray}

\noindent form an orthogonal and complete set

\begin{eqnarray} \label{orthocomp}
& & \Big( \chi^{b}(\vect{x}|\vect{k}^{\prime}),
\chi^{a}(\vect{x}|\vect{k}) \Big)
=\int \chi^{a}(\vect{x}|\vect{k}) \wedge
\overline{*} \chi^{b}(\vect{x}|\vect{k}^{\prime})
=\delta^{ab} \delta(\vect{k}-\vect{k}^{\prime}), \\ 
& & \sum_{a} \int |\chi^{a}(\vect{x}|\vect{k})>
<\chi^{a}(\vect{x}^{\prime}|\vect{k})|
d^{3}k=I\delta(\vect{x}-\vect{x}^{\prime}).
\nonumber  
\end{eqnarray}

\noindent Here we use inner product and interior 
product (bra-ket) notation for $1$-forms.

  The complex-valued Moses coframe

\begin{eqnarray} \label{coframe}
& & \omega^{a}(\vect{k}) =-\frac{\lambda}{\sqrt{2}}
\Bigg\{ \left[ \frac{k_{1}(k_{1}+i\lambda k_{2})}{k(k+k_{3})}-1 \right] dx^{1}
+\left[ \frac{k_{2}(k_{1}+i\lambda k_{2})}{k(k+k_{3})}-i\lambda \right] dx^{2}
+\frac{k_{1}+i\lambda k_{2}}{k} dx^{3} \Bigg\}, \\
& &  \hspace*{89mm} a=1: \lambda=1, a=2: \lambda=-1, \nonumber \\
& & \omega^{3}(\vect{k})
=-\frac{\tilde{k}}{k}=-\kappa, 
\hspace*{5mm}
\tilde{k}=k_{1}dx^{1}+k_{2}dx^{2}+k_{3}dx^{3}, 
\hspace*{3mm}  
k=|\vect{k}|, \nonumber
\end{eqnarray}

\noindent is dual to the basis $\{\vect{Q}_{\lambda}(\vect{k})\}$ in 
Fourier space.\cite{M1} The helicity states are given as $a=1: \lambda=1$, 
$a=2: \lambda=-1$, $a=3: \lambda=0$ with our conventions.
The basis $1$-forms of the Moses coframe respectively 
satisfy the orthogonality and completeness relations 

\begin{eqnarray} \label{orthocompfin}
& & \omega^{a}(\vect{k})
\wedge*\overline{\omega}^{b}(\vect{k})
=<\omega^{b}(\vect{k}), \omega^{a}(\vect{k})>*1, \\
& & \sum_{a} |\omega^{a}(\vect{k})>
<\omega^{a}(\vect{k})|=I,
\hspace*{5mm} I=(\delta_{ij}). \nonumber 
\end{eqnarray}

\noindent

  The coframe endows $\mathbb{R}^{3}$ with the standard metric

\begin{eqnarray}
& & ds^{2}=\eta_{ab} \omega^{a}\omega^{b} \\
& & \hspace*{6mm} =g_{ij} dx^{i}dx^{j}, \nonumber 
\end{eqnarray}

\noindent where

\begin{eqnarray} \label{nonabeliangaugetransformation}
\hspace*{8mm}
& & (\eta_{ab})= \left( \begin{array}{ccc}
0 & -1 & 0 \\
-1 & 0 & 0 \\
0 & 0 & 1
\end{array} \right), \hspace*{15mm} (g_{ij})=(\delta_{ij}).
\end{eqnarray}

 We can easily write the following relations

\begin{eqnarray} \label{f1}
& & \omega^{1}(\vect{k})=-\overline{\omega}^{2}(\vect{k}), 
\hspace*{3mm} 
\omega^{2}(\vect{k})=-\overline{\omega}^{1}(\vect{k}), 
\hspace*{3mm} 
\omega^{3}(\vect{k})=\overline{\omega}^{3}(\vect{k}), 
\nonumber \\
& & \omega^{a}(\vect{k})
=\omega^{a}(\vect{\kappa}),
\hspace*{3mm} \vect{\kappa}=\vect{k}/k, \\
& & \omega^{a}(-\vect{\kappa})
=-\frac{\kappa_{1}+i\lambda\kappa_{2}}{\kappa_{1}-i\lambda\kappa_{2}}
\,\, \overline{\omega}^{a}(\vect{\kappa}), 
\hspace*{7mm} a=1: \lambda=1, a=2: \lambda=-1, \nonumber \\
\nonumber \\
& & \omega^{3}(\vect{k}) \wedge 
\omega^{a}(\vect{k})=i\lambda *\omega^{a}(\vect{k}), 
\hspace{10mm} a=1: \lambda=1,\, a=2: \lambda=-1,\, 
a=3: \lambda=0, \nonumber \\ 
& & \omega^{a}(\vect{k}) \wedge 
\overline{\omega}^{b}(\vect{k})=i\lambda \delta^{ab}
*\omega^{3}(\vect{k}), 
\hspace{6mm} a, b=1: \lambda=1,\, \, a, b=2: \lambda=-1, \\ 
& & \omega^{3}(\vect{k}) \wedge 
*\omega^{a}(\vect{k})=0, \hspace{22mm} a=1, 2,  \nonumber \\
& & \omega^{3}(\vect{k}) \wedge 
*\omega^{3}(\vect{k})=*1, \nonumber \\
\nonumber \\
& & \omega^{1}(\vect{\kappa})
\wedge\omega^{2}(\vect{\kappa})
\wedge\omega^{3}(\vect{\kappa})=-i*1.
\end{eqnarray}

\noindent Here $(\, \bar{}\, )$ denotes complex conjugation, 
$(*)$ is the Hodge dual, $*1=dx^{1}\wedge dx^{2}\wedge dx^{3}$, 
$\omega=|\omega^{a}_{i}(\vect{\kappa})|=i$, $\sqrt{|\eta|}=i$, 
$\epsilon_{123}=1$ and $**=1$. We refer the reader to Refs. 
\onlinecite{M1, Ml1, Ml3, MD1, MD3} for a comparison with 
formulation in vectors.

\subsubsection{Trkalian field}

  The curl transform of a $1$-form field is given by an expansion 
in terms of these eigenforms as in the vectorial case. We can think 
of this as a \textit{plane wave} expansion in the coframe $\{\omega^{a}\}$.

  A Trkalian field $*F$

\begin{eqnarray} \label{fieldequation}
*d*F-\nu*F=0 ,
\end{eqnarray}

\noindent can be expressed as 

\begin{eqnarray} \label{expansion}
*F(\vect{x})
=\sum^{\hspace*{7mm} \prime}_{a} 
*\tensor*[^{a}]{F}{}(\vect{x}) ,
\end{eqnarray}

\noindent where

\begin{eqnarray} \label{helicitycomponent}
*\tensor*[^{a}]{F}{}(\vect{x})
=\frac{1}{g}
\int \chi^{a}(\vect{x}|\vect{k})
f_{a}(\vect{k}) d^{3}k,
\end{eqnarray}

\noindent excluding the divergenceful component.\cite{Ml1, Ml3, KS1} 
The factor $1/g$ is introduced for the sake of a proper strength for 
the gauge potential in topologically massive gauge theory.\cite{KS1} 
This can be taken as 1 for general Trkalian fields. Then the curl 
transform of $*\tensor*[^{a}]{F}{}(\vect{x})$ is given as 

\begin{eqnarray} \label{inverse}
& & f_{a}(\vect{k})
=g \Big( \chi^{a}(\vect{x}|\vect{k}), 
*{F}(\vect{x}) \Big).
\end{eqnarray}

  We find 

\begin{eqnarray} \label{distribution}
f_{a}(\vect{k})=\frac{\delta (k-\lambda\nu)}{k^{2}}
s_{a}(\vect{k}),
\end{eqnarray}

\noindent using a radial delta function.\cite{Ml1}
An arbitrary solution is given entirely in terms of its 
transform on the sphere of radius $k=\lambda\nu=|\nu|$. Furthermore, 
only the eigenforms for which $\lambda=sgn(\nu)$ contribute to the 
field (\ref{expansion}).\cite{Ml1} The expansion (\ref{expansion}) 
simplifies into

\begin{eqnarray} \label{simplifiedexpansion}
& & *\tensor*[^{a}]{F}{}(\vect{x})
=\frac{1}{g} \int \chi^{a}
(\vect{x}|\lambda\nu \vect{\kappa})
s_{a}(\lambda\nu \vect{\kappa}) d\Omega \\
& & \hspace*{13mm}
=\frac{1}{(2\pi)^{3/2}} \,\,  \frac{1}{g}
\int e^{i \lambda \nu \vect{\kappa} \cdot \vect{x}} 
\omega^{a} (\vect{\kappa})
s_{a}(\lambda\nu\vect{\kappa}) d\Omega, \nonumber
\end{eqnarray}

\noindent where $d\Omega$ is the spherical area element and 
$\vect{\kappa}=\vect{k}/{k}$ is a unit vector 
in transform space. Therefore a solution can be defined entirely by the 
value of its curl transform on the unit sphere in transform space.\cite{Ml1} 
We call $s_{a}$ the spherical curl transform\cite{Ml3} in order to distinguish 
it from the full curl transform $f_{a}$.

  The spherical curl transform\cite{Ml3} is given as

\begin{eqnarray} \label{sphericalcurltransform}
s_{a}(\lambda\nu\vect{\kappa})
=\frac{1}{(2\pi)^{1/2}} g\nu^{2} e^{-i\lambda\nu p} 
F^{\mathcal{R}}_{a}(p, \vect{\kappa}),
\end{eqnarray}

\noindent where

\begin{eqnarray} \label{Radoncomp}
& & F^{\mathcal{R}}_{a}(p, \vect{\kappa})
=\int *\tensor*[^{a}]{F}{}(\vect{x}) \wedge \overline{*} 
\omega^{a} (\vect{\kappa})
\delta(p-\vect{\kappa} \cdot \vect{x}) \\
& & \hspace*{15mm} 
=<\omega^{a} (\vect{\kappa}), 
\int *\tensor*[^{a}]{F}{}(\vect{x})>
\delta(p-\vect{\kappa} \cdot \vect{x}) *1,
\nonumber
\end{eqnarray}

\noindent is the Radon transform of the component $F_{a}(\vect{x})
=<\omega^{a}(\vect{\kappa}), *\tensor*[^{a}]{F}{}(\vect{x})>$ 
of $*\tensor*[^{a}]{F}{}(\vect{x})$. The Radon transform of a 
function (see Appendix \ref{RadontrfRadon}) is basically defined as the 
integral of that function over the plane at a distance 
$p=\vect{\kappa}\cdot\vect{x}$ to the origin with unit 
normal $\vect{\kappa}$. 

   We find 

\begin{eqnarray} \label{Radoncurl}
F^{\mathcal{R}}_{a}(p, \vect{\kappa})
=(2\pi)^{1/2} \, \frac{1}{g} \, \frac{1}{\nu^{2}} 
<\omega^{a}(\vect{\kappa}),
\left[ e^{i\lambda\nu p} \omega^{a}(\vect{\kappa})
s_{a}(\lambda\nu \vect{\kappa}) 
+ e^{-i\lambda\nu p} \omega^{a}(-\vect{\kappa})
s_{a}(-\lambda\nu \vect{\kappa}) \right] >, 
\nonumber
\end{eqnarray}

\noindent substituting (\ref{simplifiedexpansion}) in (\ref{Radoncomp})
and following the reasoning in Ref. \onlinecite{Ml3} with our notation. 
We shall not repeat these calculations here since they are formally 
analogous. Then we construct

\begin{eqnarray} \label{Radonform}
*\tensor*[^{a}]{F}{}^{\mathcal{R}}(p, \vect{\kappa})
=(2\pi)^{1/2} \, \frac{1}{g} \, \frac{1}{\nu^{2}} 
\left[ e^{i\lambda\nu p} \omega^{a}(\vect{\kappa})
s_{a}(\lambda\nu \vect{\kappa}) 
+ e^{-i\lambda\nu p} \omega^{a}(-\vect{\kappa})
s_{a}(-\lambda\nu \vect{\kappa}) \right],
\end{eqnarray}

\noindent using the completeness relation (\ref{orthocompfin}).
We can write the spherical curl transform (\ref{sphericalcurltransform}) 
as 

\begin{eqnarray} \label{sphericalcurltransformdifform}
& & s_{a}(\lambda\nu\vect{\kappa})
=\frac{1}{(2\pi)^{1/2}} g\nu^{2} e^{-i\lambda\nu p} 
*[ *{^{a}}{F}^{\mathcal{R}}(p, \vect{\kappa})
\wedge \overline{*} \omega^{a} (\vect{\kappa}) ] \\
& & \hspace*{13mm} =\frac{1}{(2\pi)^{1/2}} g\nu^{2} e^{-i\lambda\nu p} 
<\omega^{a}(\vect{\kappa}), 
*{^{a}}{F}^{\mathcal{R}}(p, \vect{\kappa})> \nonumber.
\end{eqnarray}

\noindent This is a Radon probe transformation\cite{NW} 
(see Appendix \ref{RadontrfRadon}). The Radon transform 
$*{^{a}}{F}^{R}(p,\vect{\kappa})$ is composed 
of both helicity components of the field.\cite{Ml3} This 
satisfies the relations 

\begin{eqnarray} \label{identities}
*\tensor*[^{a}]{F}{}^{\mathcal{R}}(-p, -\vect{\kappa})
=*\tensor*[^{a}]{F}{}^{\mathcal{R}}(p, \vect{\kappa}), 
\hspace*{8mm}
*\tensor*[^{a}]{F}{}^{\mathcal{R}}(-p, \vect{\kappa})
=*\tensor*[^{a}]{F}{}^{\mathcal{R}}(p, -\vect{\kappa}). 
\end{eqnarray}

  See Appendix \ref{SphCurlRadon} for equivalence of the 
spherical curl and Radon transformations.

   We have given a formulation of the spherical curl transformation
in terms of differential forms. We shall use vector formulation below 
which is more suitable for Radon transformation in three dimensions. 
One can use the correspondence between differential forms and vectors 
for translation.

\subsection{The Radon transformation}
\label{Radontrfs}

  The Radon transform $\vect{F}^{\mathcal{R}}_{\lambda}(\vect{\kappa}\cdot
\vect{x}, \vect{\kappa})$, (\ref{Radonform}) for  
$p=\vect{\kappa}\cdot\vect{x}$ which is the integral 
of the Trkalian field over the hyperplane orthogonal to $\vect{\kappa}$ 
containing $\vect{x}$ satisfies

\begin{eqnarray} \label{Radonfield}
\vect{\nabla}\bm{\times}\vect{F}^{\mathcal{R}}_{\lambda}
(\vect{\kappa}\cdot\vect{x}, \vect{\kappa})
-\nu\vect{F}^{\mathcal{R}}_{\lambda}
(\vect{\kappa}\cdot\vect{x}, \vect{\kappa})=0,
\end{eqnarray}

\noindent (\ref{fieldequation}), where 
$\vect{\nabla}\cdot\vect{F}^{\mathcal{R}}_{\lambda}
(\vect{\kappa}\cdot\vect{x}, \vect{\kappa})=0$. 

  We shall prove that the Radon transform of a Trkalian field 
satisfies a corresponding eigenvalue equation in the transform 
space. We shall interchangeably use $[\,\,]^{\mathcal{R}}$ or 
$\vect{\mathcal{R}}[\,\,]$ for denoting the Radon transform. 
(See Appendix \ref{RadontrfRadon}.) 
  
\begin{dfn}
The operator $\vect{\Gamma}$ is defined as

\begin{eqnarray}
\vect{\Gamma}=\vect{\kappa}\frac{\partial}{\partial p}
=\frac{\partial}{\partial p}\vect{\kappa}. \nonumber
\end{eqnarray}
\end{dfn}

\noindent Here $\partial/{\partial p}$ is called the 
(infinitesimal) parallel displacement operator of the 
plane.\cite{GGV}

\begin{prop} \label{prop1}
The Radon transform intertwines: 

\vspace*{2mm}

\noindent a) the curl operator $\vect{\nabla}\bm{\times}$ and 
$\vect{\Gamma}\bm{\times}$ 

\begin{eqnarray} \label{prop1a}
\vect{\mathcal{R}}[\vect{\nabla} \bm{\times} \vect{V}
(\vect{x})](p, \vect{\kappa})=\vect{\Gamma} 
\bm{\times} \vect{\mathcal{R}}[\vect{V}(\vect{x})]
(p, \vect{\kappa}), \nonumber
\end{eqnarray}

\noindent b) the divergence operator $\vect{\nabla}\cdot$ and 
$\vect{\Gamma}\cdot$

\begin{eqnarray}
\vect{\mathcal{R}}[\vect{\nabla} \cdot \vect{V}
(\vect{x})](p, \vect{\kappa})=\vect{\Gamma} 
\cdot \vect{\mathcal{R}}[\vect{V}(\vect{x})]
(p, \vect{\kappa}), \nonumber
\end{eqnarray}

\noindent c) the gradient operator $\vect{\nabla}$ and 
$\vect{\Gamma}$

\begin{eqnarray}
\vect{\mathcal{R}}[\vect{\nabla} f(\vect{x})]
(p, \vect{\kappa})=\vect{\Gamma} 
\vect{\mathcal{R}}[f(\vect{x})](p, \vect{\kappa}). 
\nonumber
\end{eqnarray}

\end{prop}

  {\bf{Proof:}\,} We have

\begin{eqnarray}
& & \frac{\partial}{\partial x^{i}} \delta(p-\vect{\kappa}
\cdot \vect{x})=-\kappa_{i}\delta^{\prime}(p-\vect{\kappa}
\cdot \vect{x}), \hspace*{5mm} \frac{\partial s}{\partial x^{i}}
=-\kappa_{i}, \nonumber \\
& & \hspace*{80mm} s=p-\vect{\kappa} \cdot \vect{x}, \\
& & \delta^{\prime}(p-\vect{\kappa} \cdot \vect{x})
=\frac{d}{ds}\delta(s)=\frac{\partial}{\partial p} 
\delta(p-\vect{\kappa} \cdot \vect{x}),
\hspace*{5mm} \frac{\partial s}{\partial p}=1, \nonumber
\end{eqnarray}

\noindent which yield

\begin{eqnarray} \label{Gradientdelta}
& & \vect{\nabla} \delta(p-\vect{\kappa}
\cdot \vect{x})
=-\vect{\kappa} \delta^{\prime}(p-\vect{\kappa} 
\cdot \vect{x}) \\
& & \hspace*{22mm}
=-\vect{\kappa} \frac{\partial}{\partial p} 
\delta(p-\vect{\kappa} \cdot \vect{x}). \nonumber
\end{eqnarray}

  Thus\cite{BEGHT, D} we find

\begin{eqnarray}
& & \vect{\mathcal{R}}[\frac{\partial}{\partial x^{i}} f(\vect{x})]
(p, \vect{\kappa})=\int_{D} 
\frac{\partial f(\vect{x})}{\partial x^{i}} 
\delta(p-\vect{\kappa} \cdot \vect{x}) 
d^{3} x \\
& & \hspace*{28mm} =\kappa_{i} \frac{\partial}{\partial p} 
\vect{\mathcal{R}}[f(\vect{x})](p, \vect{\kappa}). \nonumber
\end{eqnarray}

\noindent The domain $D$ of integration is the whole 
space: $D=\mathbb{R}^{3}$ unless otherwise stated. 
This leads to componentwise proofs of a, b and c.

  a) More precisely, consider a $3$-dimesional (finite) 
region $D$ bounded by a surface $S=\partial D$. We have 

\begin{eqnarray}
& & \int_{D} \vect{\nabla} \bm{\times} 
\vect{V}(\vect{x})
\delta(p-\vect{\kappa} \cdot \vect{x}) d^{3}x 
= \int_{D} \vect{\nabla} \bm{\times}
[ \vect{V}(\vect{x})
\delta(p-\vect{\kappa} \cdot \vect{x})] d^{3}x 
-\int_{D}  \vect{\nabla} 
\delta(p-\vect{\kappa} \cdot \vect{x}) \bm{\times} 
\vect{V}(\vect{x}) d^{3}x \\
& & \hspace*{46mm} = -\int_{S} \delta(p-\vect{\kappa} \cdot 
\vect{x}) \vect{V}(\vect{x}) \bm{\times} 
d\vect{s} 
-\int_{D}  \vect{\nabla} 
\delta(p-\vect{\kappa} \cdot \vect{x}) \bm{\times} 
\vect{V}(\vect{x}) d^{3}x. \nonumber
\end{eqnarray}

\noindent The first integral vanishes if we consider functions 
$\vect{V}(\vect{x})$ which vanish at infinity, as the region $D$ 
is extended to whole space. (For a finite region we could use 
vectors normal to the surface.) Thus 

\begin{eqnarray} \label{inteq}
\vect{\nabla} \bm{\times} \vect{V}(\vect{x})
\delta(p-\vect{\kappa} \cdot \vect{x})
=-\vect{\nabla} \delta(p-\vect{\kappa} \cdot 
\vect{x}) \bm{\times} \vect{V}(\vect{x}),
\end{eqnarray}

\noindent (under integral). Hence we find

\begin{eqnarray}
& & \vect{\mathcal{R}}[\vect{\nabla} \bm{\times} \vect{V}
(\vect{x})](p, \vect{\kappa})
=\int_{D} \vect{\nabla} \bm{\times} 
\vect{V}(\vect{x})
\delta(p-\vect{\kappa} \cdot \vect{x}) d^{3} x \\
& & \hspace*{31mm} =\vect{\Gamma} \bm{\times}
\vect{\mathcal{R}}[\vect{V}(\vect{x})]
(p, \vect{\kappa}), 
\nonumber
\end{eqnarray}

\noindent using (\ref{inteq}) and (\ref{Gradientdelta}). 

  The proofs of b and c are based on a similar reasoning. The 
$(\bm{\times})$ product is respectively replaced by $(\cdot)$ and 
ordinary product, with similar boundary conditions at infinity (for a 
finite region, with certain boundary conditions). See Ref. \onlinecite{NW}
for alternative proofs with different techniques. \hfill {\boldmath $\Box$} 

\begin{dfn}
The operator $\Gamma^{2}$ is defined as
$\Gamma^{2}=\vect{\Gamma}\cdot\vect{\Gamma}$.
\end{dfn}

\begin{prop}
The operator $\Gamma^{2}$ acts on $f^{\mathcal{R}}(p, \vect{\kappa})$ as

\begin{eqnarray} \label{prop2}
\Gamma^{2}f^{\mathcal{R}}(p, \vect{\kappa})
=(\partial^{2}/{\partial p^{2}})f^{\mathcal{R}}(p, \vect{\kappa}). 
\nonumber
\end{eqnarray}
\end{prop}

  {\bf{Proof:}\,} We simply have

\begin{eqnarray}
& & \Gamma^{2} f^{\mathcal{R}}(p, \vect{\kappa})
=\vect{\Gamma} \cdot [\vect{\Gamma}
f^{\mathcal{R}}(p, \vect{\kappa})] \nonumber \\
& & \hspace*{18mm} = \vect{\kappa} \frac{\partial}{\partial p}
\cdot [\vect{\kappa} \frac{\partial}{\partial p}
f^{\mathcal{R}}(p, \vect{\kappa})] \nonumber \\
& & \hspace*{18mm} =\frac{\partial}{\partial p} (\vect{\kappa}
\cdot \vect{\kappa}) \frac{\partial}{\partial p}
f^{\mathcal{R}}(p, \vect{\kappa}) \nonumber \\
& & \hspace*{18mm} = \frac{\partial^{2}}{\partial p^{2}}
f^{\mathcal{R}}(p, \vect{\kappa}), \hspace*{5mm}
\vect{\kappa} \cdot \vect{\kappa}=1. \nonumber 
\end{eqnarray}
\hfill {\boldmath $\Box$}

\noindent This operator corresponds to $\square=\Gamma^{2}$ which is 
intertwined\cite{Helgason} with Laplacian $\nabla^{2}$. This acts 
componentwise on vectors $\vect{V}^{R}(p, \vect{\kappa})$ 
analogous to the action of $\nabla^{2}$ on 
$\vect{V}(\vect{x})$.

\begin{prop} \label{Prop3}
The operator $\vect{\Gamma}$ satisfies the following identities:

\vspace*{2mm}

\noindent a)
\vspace*{-7mm}
\begin{eqnarray} \label{prop3a}
\hspace*{-133mm}
\vect{\Gamma}\bm{\times}[\vect{\Gamma} 
f^{\mathcal{R}}(p, \vect{\kappa})]=0, \nonumber
\end{eqnarray}

\noindent b)
\vspace*{-7mm}
\begin{eqnarray} \label{prop3b}
\hspace*{-129mm} 
\vect{\Gamma}\cdot[\vect{\Gamma} 
\bm{\times}\vect{V}^{\mathcal{R}}(p, \vect{\kappa})]=0, 
\nonumber
\end{eqnarray}

\noindent c)
\vspace*{-7mm}
\begin{eqnarray} \label{prop3c}
\hspace*{-83mm}
\vect{\Gamma}\bm{\times}[\vect{\Gamma} 
\bm{\times}\vect{V}^{\mathcal{R}}(p, \vect{\kappa})]
=\vect{\Gamma} [\vect{\Gamma} 
\cdot\vect{V}^{\mathcal{R}}(p, \vect{\kappa})] 
-\Gamma^{2} \vect{V}^{\mathcal{R}}(p, \vect{\kappa}). 
\nonumber
\end{eqnarray}

\end{prop}

  {\bf{Proof:}\,} These follow as

\vspace*{2mm}

\noindent a)
\vspace*{-9mm}
\begin{eqnarray} 
& & \vect{\Gamma} \bm{\times} [\vect{\Gamma} 
f^{\mathcal{R}}(p, \vect{\kappa})]
= \vect{\kappa} \frac{\partial}{\partial p}
\bm{\times} [\vect{\kappa} \frac{\partial}{\partial p}
f^{\mathcal{R}}(p, \vect{\kappa})] \nonumber \\
& & \hspace*{26mm} =\frac{\partial}{\partial p} (\vect{\kappa}
\bm{\times} \vect{\kappa}) \frac{\partial}{\partial p}
f^{\mathcal{R}}(p, \vect{\kappa}) \nonumber \\
& & \hspace*{26mm} =0, \hspace*{5mm}
\vect{\kappa} \bm{\times} \vect{\kappa}=0, \nonumber 
\hspace*{51mm}
\end{eqnarray}

\noindent b)
\vspace*{-9mm}
\begin{eqnarray}
& & \vect{\Gamma} \cdot [\vect{\Gamma} \bm{\times} 
\vect{V}^{\mathcal{R}}(p, \vect{\kappa})]
= \vect{\kappa} \frac{\partial}{\partial p} \cdot
[\vect{\kappa} \frac{\partial}{\partial p} \bm{\times}
\vect{V}^{\mathcal{R}}(p, \vect{\kappa})] \nonumber \\
& & \hspace*{29mm} =\frac{\partial^{2}}{\partial p^{2}} 
\vect{\kappa} \cdot [\vect{\kappa} 
\bm{\times} \vect{V}^{\mathcal{R}}(p, \vect{\kappa})] \nonumber \\
& & \hspace*{29mm} =0, \hspace*{5mm}
\vect{\kappa} \cdot [\vect{\kappa} \bm{\times}
\vect{V}^{\mathcal{R}}(p, \vect{\kappa})] =0, \nonumber 
\hspace*{28mm}
\end{eqnarray}
 
\noindent c)
\vspace*{-9mm}
\begin{eqnarray}
& & \vect{\Gamma} \bm{\times} [\vect{\Gamma} \bm{\times}
\vect{V}^{\mathcal{R}}(p, \vect{\kappa})]
=\vect{\kappa} \frac{\partial}{\partial p} \bm{\times}
[\vect{\kappa} \frac{\partial}{\partial p} \bm{\times}
\vect{V}^{\mathcal{R}}(p, \vect{\kappa})] \nonumber \\
& & \hspace*{31mm} =\frac{\partial^{2}}{\partial p^{2}} 
\vect{\kappa} \bm{\times} [\vect{\kappa} \bm{\times}
\vect{V}^{\mathcal{R}}(p, \vect{\kappa})] \nonumber \\
& & \hspace*{31mm} =\frac{\partial^{2}}{\partial p^{2}} 
\{[\vect{\kappa} \cdot \vect{V}^{\mathcal{R}}
(p, \vect{\kappa})] \vect{\kappa}
-(\vect{\kappa} \cdot \vect{\kappa}) 
\vect{V}^{\mathcal{R}}(p, \vect{\kappa})\} 
\nonumber \\
& & \hspace*{31mm} =\vect{\kappa}\frac{\partial}{\partial p}
[\vect{\kappa}\frac{\partial}{\partial p} \cdot 
\vect{V}^{\mathcal{R}}(p, \vect{\kappa})]
-\vect{\kappa}\frac{\partial}{\partial p} \cdot
[\vect{\kappa}\frac{\partial}{\partial p}
\vect{V}^{\mathcal{R}}(p, \vect{\kappa})] 
\nonumber \\
& & \hspace*{31mm} =\vect{\Gamma} [\vect{\Gamma} \cdot 
\vect{V}^{\mathcal{R}}(p, \vect{\kappa})] -
\Gamma^{2} \vect{V}^{\mathcal{R}}(p, \vect{\kappa}). \nonumber
\end{eqnarray}
\hfill {\boldmath $\Box$}

\begin{prop}
The Radon transform intertwines the eigenvalue 
operators $(\vect{\nabla}\bm{\times})-\nu=0$ and 
$(\vect{\Gamma}\bm{\times})-\nu=0$ for constant 
eigenvalues.
\end{prop}

  {\bf{Proof:}\,} The proof follows from the 
linearity of the Radon transformation. 
\hfill {\boldmath $\Box$}

  We can easily check the Radon transform (\ref{Radonform}) 
satisfies

\begin{eqnarray} \label{transelfdualfield}
\vect{\Gamma} \bm{\times}  
\vect{F}^{\mathcal{R}}_{\lambda} 
(p, \vect{\kappa})
-\nu\vect{F}^{\mathcal{R}}_{\lambda}
(p, \vect{\kappa})=0,
\end{eqnarray}

\noindent [in differential forms: $dp=\kappa=-\omega^{3}$, 
(\ref{coframe}) since $p=\vect{\kappa}\cdot\vect{x}$].
We also have: $\vect{\kappa}\cdot
\vect{F}^{\mathcal{R}}_{\lambda}(p, \vect{\kappa})=0$ 
which leads to $\vect{\Gamma}\cdot
\vect{F}^{\mathcal{R}}_{\lambda}(p, \vect{\kappa})=0$. 
We can write (\ref{transelfdualfield}) as

\begin{equation} 
\frac{\partial}{\partial p}
\vect{F}^{\mathcal{R}}_{\lambda} 
(p, \vect{\kappa}) 
+\nu\kappa\bm{\times} 
\vect{F}^{\mathcal{R}}_{\lambda} 
(p, \vect{\kappa}) 
=0. 
\end{equation}

  A change of sign: $p\longrightarrow-p$ or 
$\vect{\kappa}\longrightarrow-\vect{\kappa}$
(but not both) leads to a change of sign in the eigen-value: 
$\nu\longrightarrow-\nu$ in (\ref{transelfdualfield}). We shall 
discuss this again in Section \ref{tmgf}. The Radon transform 
$\vect{\Gamma}\, U^{\mathcal{R}}(p, \vect{\kappa})$ of an arbitrary 
gauge transformation $\vect{\nabla}U(\vect{x})$ is normal\cite{KS1} 
to $\mathbb{S}^{2}$. 

  We need the adjoint Radon transformation 
for discussion in the reverse direction.

\begin{dfn} \label{adjointRadon}
The adjoint\cite{Helgason, Markoe} Radon transform of a function 
$\vect{G}(p, \vect{\kappa})$ on the transform space is

\begin{eqnarray} 
\vect{G}^{{\mathcal{R}}^{\dagger}}(\vect{x})
=\vect{\mathcal{R}}^{\dagger}[\vect{G}(p, \vect{\kappa})](\vect{x})
= \int_{S^{2}} \vect{G}(\vect{\kappa} \cdot \vect{x}, \vect{\kappa}) 
d\Omega. \nonumber
\end{eqnarray}
\end{dfn}

  The function $\vect{G}(\vect{\kappa}\cdot\vect{x}, \vect{\kappa})$
is to be taken as $\vect{G}(\vect{\kappa}\cdot\vect{x}, \vect{\kappa})
=\vect{F}^{\mathcal{R}}(\vect{\kappa}\cdot\vect{x}, \vect{\kappa})$. 
The Radon transform $\vect{\mathcal{R}}$ integrates over the set of points 
in a hyperplane, the adjoint transform $\vect{\mathcal{R}}^{\dagger}$ 
integrates over the set of hyperplanes through a point.\cite{Helgason, N}

\begin{prop}
The adjoint Radon transform intertwines:

\vspace*{2mm}

\noindent a) the operators $\vect{\Gamma}\bm{\times}$ 
and curl $\vect{\nabla}\bm{\times}$ 

\begin{eqnarray}
\vect{\mathcal{R}}^{\dagger}[\vect{\Gamma}\bm{\times}
\vect{G}(p, \vect{\kappa})]
(\vect{x})
=\vect{\nabla}\bm{\times} \vect{\mathcal{R}}^{\dagger}
[\vect{G}(p, \vect{\kappa})]
(\vect{x}), \nonumber
\end{eqnarray}

\vspace*{5mm}

\noindent b) the operators $\vect{\Gamma}\cdot$ 
and divergence  $\vect{\nabla}\cdot$ 

\begin{eqnarray}
\vect{\mathcal{R}}^{\dagger}[\vect{\Gamma}\cdot
\vect{G}(p, \vect{\kappa})]
(\vect{x})
=\vect{\nabla}\cdot \vect{\mathcal{R}}^{\dagger}
[\vect{G}(p, \vect{\kappa})]
(\vect{x}), \nonumber
\end{eqnarray}

\vspace*{5mm}

\noindent c) the operators $\vect{\Gamma}$ 
and gradient $\vect{\nabla}$ 

\begin{eqnarray}
\vect{\mathcal{R}}^{\dagger}[\vect{\Gamma}
\phi(p, \vect{\kappa})]
(\vect{x})
=\vect{\nabla} \vect{\mathcal{R}}^{\dagger}
[\phi(p, \vect{\kappa})]
(\vect{x}). \nonumber
\end{eqnarray}

\end{prop}

  {\bf{Proof:}\,} The componentwise proofs 
of a, b and c follow from the identity

\begin{eqnarray}
& & \frac{\partial}{\partial x^{i}}
\vect{\mathcal{R}}^{\dagger}[\phi(p, \vect{\kappa})]
(\vect{x})
=\frac{\partial}{\partial x^{i}} \int_{S^{2}}
\phi(\vect{\kappa}\cdot\vect{x}, 
\vect{\kappa})d\Omega \\
& & \hspace*{29mm} = \int_{S^{2}} \kappa^{i} 
\frac{\partial}{\partial p}
\phi(p, \vect{\kappa})d\Omega,
\hspace*{3mm} 
p=\vect{\kappa}\cdot\vect{x} \nonumber \\
& & \hspace*{29mm} =\vect{\mathcal{R}}^{\dagger}
[\kappa^{i} \frac{\partial}{\partial p}
\phi(p, \vect{\kappa})](\vect{x}). \nonumber
\end{eqnarray}
\hfill {\boldmath $\Box$}

\noindent The adjoint Radon transform also intertwines\cite{Helgason} 
the operators $\Gamma^{2}$ and $\nabla^{2}$. 

\begin{prop} The adjoint Radon transform intertwines the 
eigenvalue operators $(\vect{\Gamma}\bm{\times})-\nu=0$ 
and $(\vect{\nabla}\bm{\times})-\nu=0$ for constant eigenvalues.
\end{prop}

  {\bf{Proof:}\,} The proof follows from the linearity 
of the adjoint Radon transformation. 
\hfill {\boldmath $\Box$}

  Thus, we have a correspondence between the (constant) eigenvalue 
equations $(\vect{\nabla}\bm{\times})-\nu=0$ in the physical space 
and $(\vect{\Gamma}\bm{\times})-\nu=0$ in the transform space. Note 
that we should employ $(-1/8\pi^{2})\nabla^{2}$ on 
$\vect{\mathcal{R}}^{\dagger}[\vect{F}^{\mathcal{R}}
(p, \vect{\kappa})]$ for a complete inversion. This simplifies for 
a Trkalian field. Further one should consider the appropriate class 
of functions for the Radon transform and its adjoint.\cite{Helgason} 
(See Appendix \ref{RadontrfRadon}.)

\subsubsection{A refinement}  
\label{Ref}

  The Radon transform (\ref{Radonform}) of a Trkalian field is composed 
of both helicity components of the field. The inverse Radon transform 
(\ref{inverseRadon}) requires knowledge of the transform evaluated over 
the entire sphere.\cite{MP} However the knowledge of a helicity component 
evaluated over the entire sphere suffices to reproduce the knowledge of the 
other component. Thus construction of the field can be achieved using the 
knowledge of a single component evaluated over the entire sphere as in 
(\ref{simplifiedexpansion}). Equivalently, the field can be constructed 
using knowledge of both components, i. e. the Radon transform 
(\ref{Radonform}), on a canonical hemisphere which can consist 
of disconnected parts. For scalar fields, a similar result is proved 
in Ref. \onlinecite{MP}. Here, this easily follows as one considers both 
helicity components in the Radon transform of a Trkalian field. This also 
removes redundancy of double-covering of space.\cite{MP}

  A canonical hemisphere $H$ is a (Lebesgue measurable) subset of 
the unit $2$-sphere $\mathbb{S}^{2}$ whose area is $2\pi$ such that 
if the tip of a unit vector $\vect{\kappa}$ based at the origin is 
in the hemisphere then the tip of $-\vect{\kappa}$ is not in the 
hemispere.\cite{MP} The complementary set $H^{\prime}$ in $\mathbb{S}^{2}$ 
is a canonical hemisphere too. A canonical hemisphere can consists of 
disconnected parts. 

   If we substitute (\ref{Radonform}) with
$p=\vect{\kappa}\cdot\vect{x}$ in (\ref{inverseRadon}), 
we find

\begin{eqnarray}  \label{refinement2}
& & \vect{F}_{\lambda}(\vect{x})
=-\frac{1}{(2\pi)^{3/2}} \, \frac{1}{g} \, \frac{1}{\nu^{2}} 
\nabla^{2}_{x} \int_{H} \left[ e^{i\lambda\nu \vect{\kappa} \cdot \vect{x}} 
\vect{Q}_{\lambda}(\vect{\kappa})s_{\lambda}(\lambda\nu \vect{\kappa}) 
+ e^{-i\lambda\nu \vect{\kappa} \cdot \vect{x}} \vect{Q}_{\lambda}(-\vect{\kappa})
s_{\lambda}(-\lambda\nu \vect{\kappa}) \right] d\Omega \\
& & \hspace*{11mm} =\frac{1}{(2\pi)^{3/2}} \, \frac{1}{g} 
\int_{H} \left[ e^{i\lambda\nu \vect{\kappa}\cdot\vect{x}} 
\vect{Q}_{\lambda}(\vect{\kappa})s_{a}(\lambda\nu \vect{\kappa}) 
+ e^{-i\lambda\nu \vect{\kappa} \cdot \vect{x}} \vect{Q}{\lambda}(-\vect{\kappa})
s_{\lambda}(-\lambda\nu \vect{\kappa}) \right] d\Omega, \nonumber
\end{eqnarray}

\noindent as $\vect{\kappa} \longrightarrow -\vect{\kappa}$ in $H^{\prime}$, 
using the identities (\ref{identities}). This can also be inferred from 
(\ref{simplifiedexpansion}).

  We can see this on a simple example choosing 
$s_{\lambda}(\lambda\nu \vect{\kappa})=(2\pi)^{3/2} 
g\delta( \vect{\kappa}-\vect{\kappa}_{0} )$.
We decompose the sphere $\mathbb{S}^{2}$ into two complementary 
canonical hemispheres $H$ and $H^{\prime}$ where the tip of 
$\vect{\kappa}_{0}$ is contained in $H$:
$\vect{\kappa}_{0} \in H$. If we take
$s_{\lambda}(-\lambda\nu \vect{\kappa})=(2\pi)^{3/2} g
\delta( \vect{\kappa}-\vect{\kappa}^{\prime}_{0} )$,
($\vect{\kappa} \longrightarrow -\vect{\kappa}$: 
$\vect{\kappa}_{0} \longrightarrow 
\vect{\kappa}^{\prime}_{0}= -\vect{\kappa}_{0}$)
then $\vect{\kappa}^{\prime}_{0} \in H^{\prime}$. We find 

\begin{eqnarray} \label{exampleinverseRadon}
& & \vect{F}_{\lambda}(\vect{x})
=e^{i\lambda\nu \vect{\kappa}_{0}\cdot\vect{x}} 
\vect{Q}_{\lambda}(\vect{\kappa}_{0}) \\
& & \hspace*{11mm} 
=e^{-i\lambda\nu \vect{\kappa}^{\prime}_{0}\cdot\vect{x}} 
\vect{Q}_{\lambda}(-\vect{\kappa}^{\prime}_{0}), \nonumber
\end{eqnarray}

\noindent using (\ref{refinement2}) or equivalently 
inverting it and integrating over $H^{\prime}$. 

It is possible to define a left and right inverse of this refined 
(inverse) transformation if it is applied to Radon transforms of functions 
(see Appendix \ref{RadontrfRefinement}). Alternatively, one can introduce 
inverse of this refined transformation redefining the Radon transform and 
the set $\{\vect{F}^{\mathcal{R}}(p, \vect{\kappa})\}$ to require that 
$\vect{\kappa}$ be restricted into a canonical hemisphere. Then 
$\vect{\mathcal{R}}^{-1}_{H}$ is both a left and right inverse of 
$\vect{\mathcal{R}}_{H} $: 
$\vect{\mathcal{R}}^{-1}_{H}\vect{\mathcal{R}}_{H}[\vect{F}](\vect{x})
=\vect{F}(\vect{x})$, $\vect{\mathcal{R}}_{H}\vect{\mathcal{R}}^{-1}_{H}
[\vect{F}^{\mathcal{R}}](p, \vect{\kappa})=\vect{F}^{\mathcal{R}}
(p, \vect{\kappa})$.\cite{MP} 

\section{RADON TRANSFORM AND BIOT-SAVART INTEGRAL}
\label{RBS}

  In this section, we shall consider the relation of the Radon 
transformation with Riesz potential and $BS$ integral. We shall 
also present an Ampere law type relation for Trkalian fields. 
Then we shall discuss Radon transform of the $BS$ integral.

  We can easily\cite{D, Helgason, Markoe, N} show 

\begin{eqnarray} \label{adjointRadonRadon}
\vect{\mathcal{R}}^{\dagger}\vect{\mathcal{R}}[\vect{F}]
(\vect{x})
=8\pi^{2} I^{2}[\vect{F}](\vect{x}).
\end{eqnarray}

\noindent The integral on the right-hand side which is 
defined as the convolution of $\vect{F}$ with Riesz kernel 
$K_{2}=(1/4\pi)|\vect{x}|^{-1}$

\begin{eqnarray} \label{Rieszpotential}
& & I^{2}[\vect{F}](\vect{x})
=(K_{2}*\vect{F})(\vect{x}) \\
& & \hspace*{15mm} =\frac{1}{4\pi}\int_{D}\frac{\vect{F}(\vect{y})}
{|\vect{x}-\vect{y}|} d^{3}y, \nonumber
\end{eqnarray}

\noindent is called the Riesz potential for the vector 
field $\vect{F}(\vect{x})$. We remind: $D=\mathbb{R}^{3}$ 
for the Radon transformation. The operator $-\nabla^{2}$ 
is an (left) inverse of the Riesz integral.\cite{N} 

  We find

\begin{eqnarray} 
\vect{\nabla}_{x} \bm{\times} 
\vect{\mathcal{R}}^{\dagger}\vect{\mathcal{R}}[\vect{F}]
(\vect{x})
=8\pi^{2} BS[\vect{F}](\vect{x}),
\end{eqnarray}

\noindent where 

\begin{eqnarray} \label{BiotSavart}
& & BS[\vect{F}](\vect{x})
=\vect{\nabla}_{x}\bm{\times}
I^{2}[\vect{F}](\vect{x}) \\
& & \hspace*{17mm}=\frac{1}{4\pi}\int_{D}
\vect{F}(\vect{y}) 
\bm{\times} \frac{\vect{x}-\vect{y}}
{|\vect{x}-\vect{y}|^{3}}
d^{3}y, \nonumber
\end{eqnarray}

\noindent is the $BS$ integral operator.\cite{CDG} The 
\textit{vector potential} for $BS[\vect{F}](\vect{x})$ 
is given by the Riesz potential (\ref{Rieszpotential}).\cite{CDG}

  We immediately see that

\begin{eqnarray}
\vect{\nabla}_{x} \cdot \vect{\nabla}_{x}\bm{\times}
\vect{\mathcal{R}}^{\dagger}\vect{\mathcal{R}}[\vect{F}]
(\vect{x})
=8\pi^{2} \vect{\nabla}_{x} \cdot 
BS[\vect{F}](\vect{x})=0.
\end{eqnarray}

\noindent We also find

\begin{eqnarray} \label{RadonCurl}
\vect{\nabla}_{x} \bm{\times} \vect{\nabla}_{x} 
\bm{\times} \vect{\mathcal{R}}^{\dagger}\vect{\mathcal{R}}[\vect{F}]
(\vect{x})
=8\pi^{2} \vect{\nabla}_{x} \bm{\times}
BS[\vect{F}](\vect{x})
=8\pi^{2} \vect{F}(\vect{x}),
\end{eqnarray}

\noindent if $\vect{F}$ is divergence-free and it also vanishes 
at infinity (for a finite region $D$, it is tangent to the surface 
$S=\partial D$ bounding this region). (See Appendix \ref{BS}.) Hence 
the $BS$ integral is divergence-free and the curl operator is 
a left inverse of this under these conditions.\cite{CDG, P}

  We can associate an eigenvalue equation for $BS$ operator (the 
eigenvalue being reciprocal of $\nu$) with the Trkalian fields only 
in this case.\cite{CDG, P} In general, the curl can be taken to be 
the inverse of an appropriately modified $BS$ operator.\cite{CDGT1,CDGT2}

  As for the Trkalian fields, we find

\begin{eqnarray} \label{RadontopmasRieszBiotSavart}
& & \vect{\mathcal{R}}^{\dagger}\vect{\mathcal{R}}
[\vect{F}_{\lambda}](\vect{x})
=8\pi^{2} \frac{1}{\nu^{2}} \vect{F}_{\lambda}(\vect{x}), 
\end{eqnarray}

\noindent using Definition \ref{adjointRadon} and (\ref{Radonform}),
(\ref{simplifiedexpansion}). This can be inferred from (\ref{inverseRadon}) 
using $\Gamma^{2}$ which is intertwined with $\nabla^{2}$. Note that the 
conditions in (\ref{RadonCurl}) are satisfied for a Trkalian field if its 
Radon transform exists (see Appendix \ref{RadontrfRadon}). The equation 
(\ref{RadontopmasRieszBiotSavart}) yields

\begin{eqnarray} \label{TopmassInv}
\vect{\nabla}_{x} \bm{\times} \vect{\nabla}_{x}\bm{\times}
\vect{\mathcal{R}}^{\dagger}\vect{\mathcal{R}}
[\vect{F}_{\lambda}](\vect{x})
=8\pi^{2}\vect{F}_{\lambda}(\vect{x}).
\end{eqnarray}

\noindent This reduces to the inverse Radon transform 
(\ref{inverseRadon})  using (\ref{Radonfield}) on the 
left-hand side. 

  We find an Ampere law type relation\cite{KS2, KS3}

\begin{eqnarray} \label{charge}
\Phi=\nu Q,
\end{eqnarray}

\noindent integrating the equation
(\ref{fieldequation}) over a surface $S$ 
bounded by the curve $C=\partial S$. Here 

\begin{eqnarray} \label{magneticcharge}
Q=\int_{S} \vect{F}(\vect{x})
\cdot d\vect{s},
\hspace*{10mm}
\Phi=\int_{S} [\vect{\nabla} \bm{\times} 
\vect{F}(\vect{x})]
\cdot d\vect{s},
\end{eqnarray}

\noindent are respectively the flux of $\vect{F}(\vect{x})$ 
and $\vect{\nabla}\bm{\times}\vect{F}(\vect{x})$ through the 
surface $S$. The flux $\Phi$ reduces to 

\begin{eqnarray} \label{electriccirculation}
\Phi=\int_{C} \vect{F}(\vect{x})
\cdot  d\vect{l},
\end{eqnarray}

\noindent the circulation of $\vect{F}(\vect{x})$ on $C$. Thus 
the flux of a Trkalian field through a surface is determined by 
its circulation on the boundary of the surface.\cite{KS2, KS3} 
Note that the equation (\ref{Radonfield}) yields an analogous relation 
for $\vect{F}^{\mathcal{R}}_{\lambda}(\vect{\kappa}\cdot\vect{x}, \vect{\kappa})$.

\subsection{Radon transform of the Biot-Savart integral}

  We shall make use of Fourier slice theorem\cite{Helgason, Markoe}

\begin{eqnarray} \label{Fourierslice}
\vect{{\mathcal{F}}}[\vect{F}^{\mathcal{R}}(p, \vect{\kappa})]
(k, \vect{\kappa})
=2\pi\vect{{\mathcal{F}}}[\vect{F}(\vect{x})](k\vect{\kappa}),
\end{eqnarray}

\noindent for finding Radon transform of the $BS$ integral. On the 
left-hand side $\vect{{\mathcal{F}}}$ stands for $1$-dimensional 
Fourier transform in the first variable whereas on the right-hand 
side it is a vector Fourier transform in $3$ dimensions. We also 
need Fourier transform of the Riesz\cite{Helgason, Markoe} integral

\begin{eqnarray} \label{FourierRiesz}
& & \vect{{\mathcal{F}}}\{I^{2}[\vect{F}](\vect{x})\}(\vect{k})
=\frac{1}{k^{2}}\vect{{\mathcal{F}}}[\vect{F}(\vect{x})](k\vect{\kappa}) \\
& & \hspace*{26mm} =\frac{1}{2\pi} \frac{1}{k^{2}}
\vect{{\mathcal{F}}}[\vect{F}^{\mathcal{R}}(q, \vect{\kappa})](k, \vect{\kappa}).
\nonumber
\end{eqnarray}

\noindent Here the last line follows from (\ref{Fourierslice}). 

  We find 

\begin{eqnarray}
& & \vect{{\mathcal{F}}}\{I^{2}[\vect{F}]^{\mathcal{R}}(p, \vect{\kappa})\} 
(k, \vect{\kappa}) 
=2\pi \vect{{\mathcal{F}}}\{I^{2}[\vect{F}](\vect{x})\}
(k\vect{\kappa}) \\
& & \hspace{36mm} = \frac{1}{k^{2}} 
\vect{{\mathcal{F}}}[\vect{F}^{\mathcal{R}}(q, \vect{\kappa})]
(k, \vect{\kappa}), \nonumber 
\end{eqnarray}

\noindent using (\ref{Fourierslice}) and (\ref{FourierRiesz}). 
This yields Radon transform of the Riesz integral

\begin{eqnarray} \label{RadonRiesz}
& & \vect{\mathcal{R}}\{I^{2}[\vect{F}](\vect{x})\}(p, \vect{\kappa}) 
=\vect{{\mathcal{F}}}^{-1}
\{\frac{1}{k^{2}}\vect{{\mathcal{F}}}
[\vect{F}^{\mathcal{R}}(q, \vect{\kappa})]
(k, \vect{\kappa})\} (p, \vect{\kappa}) \\
& & \hspace*{30mm} =\frac{1}{2\pi}\int
\left[ \int e^{ik(p-q)}\frac{1}{k^{2}}dk \right]
\vect{F}^{R}(q, \vect{\kappa})dq.
\nonumber
\end{eqnarray}

\noindent This equivalently corresponds to
$\vect{\mathcal{R}}\vect{\mathcal{R}}^{\dagger}\vect{\mathcal{R}}
[\vect{F}](\vect{x})$, (\ref{adjointRadonRadon}).\cite{RK} This is 
a convolution for the Radon transformation.\cite{Helgason, Markoe} 
The operator $-\Gamma^{2}$ is an (left) inverse of the Radon 
transform of the Riesz integral. We define Radon-Biot-Savart 
integral operator as

\begin{eqnarray} \label{RadonBiotSavart}
& & RBS[\vect{F}^{\mathcal{R}}(q, \vect{\kappa})](p, \vect{\kappa})
=BS[\vect{F}]^{\mathcal{R}} (p, \vect{\kappa}) \\
& & \hspace*{33mm} =\vect{\Gamma}_{p}\bm{\times}
\vect{{\mathcal{F}}}^{-1}\{\frac{1}{k^{2}}\vect{{\mathcal{F}}}
[\vect{F}^{\mathcal{R}}(q, \vect{\kappa})](k, \vect{\kappa})\} 
(p, \vect{\kappa}) \nonumber \\
& & \hspace*{33mm} =\vect{{\mathcal{F}}}^{-1}\{\frac{1}{k^{2}}
\vect{{\mathcal{F}}}[\vect{\Gamma}_{q}\bm{\times}\vect{F}^{\mathcal{R}}
(q, \vect{\kappa})](k, \vect{\kappa})\} 
(p, \vect{\kappa}),
\nonumber
\end{eqnarray}

\noindent the Radon transform of the $BS$ integral
using Proposition \ref{prop1} and (\ref{BiotSavart}), 
(\ref{RadonRiesz}).

   We immediately see that 

\begin{eqnarray}
\vect{\Gamma}\cdot RBS[\vect{F}^{\mathcal{R}}
(q, \vect{\kappa})] (p, \vect{\kappa})=0,
\end{eqnarray}

\noindent using Proposition \ref{Prop3}. We find

\begin{eqnarray}
& & \vect{\Gamma}\bm{\times} RBS[\vect{F}^{\mathcal{R}}
(q, \vect{\kappa})] (p, \vect{\kappa})
=-\vect{\kappa}\bm{\times}[\vect{\kappa}\bm{\times}
\vect{F}^{\mathcal{R}}
(p, \vect{\kappa})] \\
& & \hspace*{40mm} =-\{[\vect{\kappa}\cdot 
\vect{F}^{\mathcal{R}}(p, \vect{\kappa})] \vect{\kappa}
-(\vect{\kappa}\cdot\vect{\kappa})
\vect{F}^{\mathcal{R}}(p, \vect{\kappa}) \}. \nonumber 
\end{eqnarray}

\noindent This yields

\begin{eqnarray}
\vect{\Gamma}\bm{\times} RBS[\vect{F}^{\mathcal{R}}
(q, \vect{\kappa})] (p, \vect{\kappa})
=\vect{F}^{\mathcal{R}}(p, \vect{\kappa}), 
\end{eqnarray}

\noindent if $\vect{\kappa}\cdot\vect{F}^{\mathcal{R}}=0$. This condition
leads to $\vect{\Gamma}\cdot\vect{F}^{\mathcal{R}}=0$ [as can be inferred 
from (\ref{RadonBiotSavart})] which corresponds to $\vect{\nabla}
\cdot\vect{F}=0$ in the physical space. Thus we conclude that 
$\vect{\Gamma}\bm{\times}$ is a left inverse of the $RBS$ operator, if 
$\vect{\kappa}\cdot\vect{F}^{\mathcal{R}}=0$. We can associate an eigenvalue 
equation for $RBS$ operator with the Radon transform of Trkalian fields. 

  The $RBS$ operator also acts as a left inverse

\begin{eqnarray}
RBS[\vect{\Gamma}_{q}\bm{\times}\vect{F}^{\mathcal{R}}(q, \vect{\kappa})] 
(p, \vect{\kappa})
=\vect{F}^{\mathcal{R}}(p, \vect{\kappa}),
\end{eqnarray}

\noindent of the operator $\vect{\Gamma}\bm{\times}$ 
for those vector fields $\vect{F}^{\mathcal{R}}$ with 
$\vect{\Gamma}\cdot\vect{F}^{\mathcal{R}}=0$. 
For example, the Radon transform (\ref{Radonform}) 
of a Trkalian field is an eigenvector

\begin{eqnarray} \label{RadonBiotSavartTopmas}
RBS[\vect{F}^{\mathcal{R}}_{\lambda}(q, \vect{\kappa})]
(p, \vect{\kappa})=\frac{1}{\nu}
\vect{F}^{\mathcal{R}}_{\lambda}(p, \vect{\kappa}),
\end{eqnarray}

\noindent of the $RBS$ operator with reciprocal eigenvalue. 
The kernel of the $RBS$ operator consists of those vector fields 
for which $\vect{\Gamma}\bm{\times}\vect{F}^{\mathcal{R}}=0$ i. e. 
\textit{gauge terms} $\vect{\Gamma}\, U^{\mathcal{R}}$, (Proposition 
\ref{Prop3}) corresponding to gradient fields $\vect{\nabla} U$ in 
the physical space.\cite{CDG}

\section{EXAMPLES}
\label{Sol}

  In this section we shall first present applications of our 
constructions on the L solution. Then we shall briefly discuss 
the CK method with circular and elliptic cylindrical solutions. 
We shall make use of an analogous method for finding solution 
of the corresponding equation in the transform space.

\subsection{Lundquist solution}

  The L\cite{L} solution is 

\begin{eqnarray} \label{lundquist}
\vect{F}_{L}=F_{0}[J_{1}(\nu r)\vect{e}_{\theta} 
+J_{0}(\nu r)\vect{e}_{z}],
\end{eqnarray}

\noindent where $J_{m}$ is the Bessel function of order $m$ and $F_{0}$ 
is the strength of the field on $z$-axis. Note this is of helicity 
$\lambda=1$.\cite{Ml3} The spherical curl transform 
(\ref{sphericalcurltransform}) of (\ref{lundquist}) 
is given by

\begin{eqnarray}
s_{1}(\nu\vect{\kappa})
=-2^{1/2} (2\pi)^{1/2} g F_{0}\, \delta(\kappa_{z})
e^{-i\psi}.
\end{eqnarray}

\noindent The radial degree of freedom $k$ is also taken into 
account.\cite{Ml3} Here: $\vect{k}=k_{r}\cos\psi\vect{e}_{x}
+k_{r}\sin\psi\vect{e}_{y}+k_{z}\vect{e}_{z}$ in the circular 
cylindrical coordinates in transform space. Thus the spherical 
curl transform of the L field lives on the equatorial circle 
of the sphere.\cite{Ml3} 

  This solution provides a simple example of Ampere law. Consider 
a disc $S$ which is bounded by the circle $C=\partial S$ of radius 
$R$, centered at the origin in $xy$-plane. We find

\begin{eqnarray}
\Phi=\nu Q=2\pi F_{0} \, RJ_{1}(\nu R), 
\end{eqnarray}

\noindent (\ref{charge}). Note that 
$\Phi=Q=0$ at zeros of $J_{1}(\nu R)$. 

  The L field is an eigenvector of the $BS$ operator

\begin{eqnarray}
BS[\vect{F}_{L}](\vect{x})
= \frac{1}{\nu} \vect{F}_{L}(\vect{x}),
\end{eqnarray}

\noindent (\ref{BiotSavart}). This calculation 
is given in Appendix \ref{BSL}.

  We find

\begin{eqnarray} \label{RadonLundquist}
\vect{F}^{\mathcal{R}}_{L}(p, \vect{\kappa}) 
=2\pi iF_{0}\frac{1}{\nu^{2}} \delta(\kappa_{z}) (e^{i\nu p}  
\vect{L}+e^{-i\nu p} \vect{L}^{\prime}), 
\end{eqnarray}

\noindent where $\vect{L}=\sin\psi\vect{e}_{x}
-\cos\psi\vect{e}_{y}-i\vect{e}_{z}$ 
and $\vect{L}^{\prime}=-\sin\psi\vect{e}_{x}
+\cos\psi\vect{e}_{y}-i\vect{e}_{z}$.
Note that we have taken the cases: $k$ negative or positive 
into account here. We immediately see that 
$\vect{F}^{\mathcal{R}}_{L}(p, \vect{\kappa})$ 
satisfies the equation (\ref{transelfdualfield}) anticipating the 
effect of $\delta(\kappa_{z})$.\cite{Ml3} This is also an 
eigenvector (\ref{RadonBiotSavartTopmas}) of the $RBS$ operator. 

\subsection{Chandrasekhar-Kendall cylindrical solutions}

   Chandrasekhar-Kendall\cite{CK} developed a method for deriving 
solutions to $\vect{\nabla}\bm{\times}\vect{F}=\nu\vect{F}$ from 
the scalar Helmholtz equation. See also Refs. \onlinecite{M, B1, B2}. 

  We find  

\begin{eqnarray} \label{Helmholtzvector}
(\nabla^{2} + \nu^{2}) \vect{F}=0,
\end{eqnarray}

\noindent applying curl on $\vect{\nabla}\bm{\times}\vect{F}
-\nu\, \vect{F}=0$. Thus every solution of $\vect{\nabla}\bm{\times}
\vect{F}-\nu\, \vect{F}=0$ is a solution of (\ref{Helmholtzvector}), 
although the converse is not necessarily true. If $\Psi$ is a scalar 
function satisfying the Helmholtz equation

\begin{eqnarray}
(\nabla^{2} + \nu^{2}) \Psi=0,
\end{eqnarray}

\noindent then three independent solutions\cite{CK} 
of the equation (\ref{Helmholtzvector}) are

\begin{eqnarray}
\vect{p}=\vect{\nabla}\bm{\times}(\Psi\vect{\omega}), \hspace*{10mm}
\vect{q}=\frac{1}{\nu}\vect{\nabla}\bm{\times}\vect{p}=\frac{1}{\nu}
\vect{\nabla}\bm{\times}\vect{\nabla}\bm{\times}(\Psi\vect{\omega}),
\hspace*{10mm} \vect{r}=\vect{\nabla}\Psi.
\end{eqnarray}

\noindent Here $\vect{\omega}$ is a fixed vector (of unit norm). 
The vectors of type $\vect{p}$ and $\vect{q}$ are respectively called toroidal 
and poloidal.\cite{LS, C, C1} We also have: $\vect{\nabla}\bm{\times}\vect{q}
=\nu\vect{p}$. Hence we find: $\vect{\nabla}\bm{\times}(\vect{p}+\vect{q})
=\nu (\vect{p}+\vect{q})$. Therefore, the most general\cite{CK, GM} solution 
of $\vect{\nabla}\bm{\times}\vect{F}-\nu\, \vect{F}=0$ among the solutions 
of (\ref{Helmholtzvector}) is 

\begin{eqnarray} \label{ChadrasekharKendall}
\vect{F}=\vect{p}+\vect{q}=\vect{\nabla}\bm{\times}
(\Psi\vect{\omega})+\frac{1}{\nu}\vect{\nabla}\bm{\times}
\vect{\nabla}\bm{\times}(\Psi\vect{\omega}).
\end{eqnarray}

\noindent The scalar $\Psi$ is called Debye potential.\cite{TdC} 
This method has been frequently used for discussing circular cylindrical, 
spherical and other type solutions.\cite{GM} For the relation with Wu-Yang 
type solutions in the topologically massive Abelian gauge theory see Refs.
\onlinecite{KS2, KS3}. 

  The CK\cite{CK} solution in cylindrical coordinates is given by 

\begin{eqnarray} \label{ChadrasekharKendallcylindrical}
\vect{F}=-[\sigma \vect{\nabla}\bm{\times}(\Psi \vect{e}_{z}) 
+ \vect{\nabla}\bm{\times}\vect{\nabla}\bm{\times}(\Psi \vect{e}_{z})],
\end{eqnarray}

\noindent where $\vect{\nabla}\bm{\times}\vect{F}-\sigma\vect{F}=0$. 
In the circular\cite{Y, TC} cylindrical coordinates 

\begin{eqnarray} \label{ChadrasekharKendallcircular}
\Psi(r, \theta, z)=J_{m}(\nu r) e^{im\theta-ikz}, 
\hspace*{10mm} \sigma^{2}=\nu^{2}+k^{2}.
\end{eqnarray}

\noindent This reduces to the L solution (\ref{lundquist}) for $m=0$, $k=0$, 
($F_{0}=-\nu^{2}$). In the elliptic cylindrical coordinates the solution is 
given by (\ref{ChadrasekharKendallcylindrical}) with

\begin{eqnarray} \label{ChadrasekharKendallelliptic}
\Psi(u, v, z)=U(u)V(v)e^{-ikz}, \hspace*{10mm} 
\sigma^{2}=k^{2}+m^{2}=\nu^{2}.
\end{eqnarray}

\noindent Here $U(u)$ and $V(v)$ are respectively the modified Mathieu 
function and the Mathieu function. For $k=0$, $\sigma=m=\nu$ and this 
solution reduces to Vandas-Romashets\cite{VR, VR1} (VR) solution which 
is an elliptic generalization of L. 

\subsubsection{Chandrasekhar-Kendall method in transform space}

  We can adapt the CK method in the transform space for finding 
solutions to $\vect{\Gamma}\bm{\times}\vect{G}=\nu\, \vect{G}$. 
Briefly, the solution of 

\begin{eqnarray} \label{Radonfieldequation}
\vect{\Gamma}\bm{\times}\vect{G}(p, \vect{\kappa})
-\nu\, \vect{G}(p, \vect{\kappa})=0,
\end{eqnarray}

\noindent that is to be found among the solutions of

\begin{eqnarray} \label{transformvectorHelmholtz}
(\Gamma^{2}+\nu^{2})\vect{G}(p, \vect{\kappa})=0,
\end{eqnarray}

\noindent is

\begin{eqnarray} \label{Potentialequation}
\vect{G}(p, \vect{\kappa})
=\vect{\Gamma}\bm{\times}(\Psi\vect{\omega})
+\frac{1}{\nu}\vect{\Gamma}\bm{\times}\vect{\Gamma}\bm{\times}
(\Psi\vect{\omega}).
\end{eqnarray}

\noindent Here $\Psi(p, \vect{\kappa})$ satisfy

\begin{eqnarray} \label{oscillator}
(\Gamma^{2}+\nu^{2})\Psi=0,
\end{eqnarray}

\noindent and $\vect{\omega}$ is a fixed vector in the transform 
space. We have ignored the superscript $R$ for Radon transform 
in $\Psi\vect{\omega}$. We remark a few important issues. First, 
the Radon transform is defined on $\mathbb{S}^{2}\bm{\times}\mathbb{R}$ 
[or $(\mathbb{R}^{3}-\{0\})\bm{\times}\mathbb{R}$], (see appendix 
\ref{RadontrfRadon}). Secondly, $\vect{G}(p, \vect{\kappa})$ may 
consist of distributions. The third issue is imposing boundary 
conditions in the transform space. 

  The first term in (\ref{Potentialequation}): 
$\vect{G}(p, \vect{\kappa})=\vect{\Gamma}\bm{\times}
(\Psi\vect{\omega})$ with $\Psi=[(2\pi)^{2}/\nu^{3}]
[e^{i\lambda\nu p} \delta(\vect{\kappa}-\vect{\kappa}_{0})
+e^{-i\lambda\nu p} \delta(\vect{\kappa}+\vect{\kappa}_{0})]$ 
and $\vect{\omega}=\vect{Q}_{\lambda}(\vect{\kappa}_{0})$ 
immediately yields

\begin{eqnarray} \label{exampleoppositehemispherefirst}
\vect{G}(p, \vect{\kappa})
=(2\pi)^{2} \, \frac{1}{\nu^{2}} 
\left[ e^{i\lambda\nu p} 
\delta(\vect{\kappa}-\vect{\kappa}_{0})
+e^{-i\lambda\nu p} 
\delta(\vect{\kappa}+\vect{\kappa}_{0}) \right]
\vect{Q}_{\lambda}(\vect{\kappa}_{0}),
\end{eqnarray}

\noindent which satisfies (\ref{Radonfieldequation}). 
This is the Radon transform:
$\vect{G}(p, \vect{\kappa})=\vect{F}^{\mathcal{R}}(p, \vect{\kappa})$  
of the solution
$\vect{F}(\vect{x})=e^{i\lambda\nu\vect{\kappa}_{0}\cdot\vect{x}}        
\vect{Q}_{\lambda}(\vect{\kappa}_{0})$.\cite{Ml3, KS1}

  As another example, consider the Radon transform 
(\ref{RadonLundquist}) of the L field. We can reproduce 
$\vect{G}(p, \vect{\kappa})=\vect{F}^{\mathcal{R}}_{L}(p, \vect{\kappa})$
from the first term in (\ref{Potentialequation}):
$\vect{G}(p, \vect{\kappa})
=\vect{\Gamma}\bm{\times}(\Psi_{1}\vect{\omega}_{1}
+\Psi_{2}\vect{\omega}_{2})$ choosing
$\Psi_{1}=(2\pi iF_{0}/\nu^{3}) \delta(\kappa_{z}) e^{i\nu p}$,  
$\Psi_{2}=(2\pi iF_{0}/\nu^{3}) \delta(\kappa_{z}) e^{-i\nu p}$ 
and $\vect{\omega}_{1}=\vect{L}$, $\vect{\omega}_{2}=\vect{L}^{\prime}$.   
The Radon transform of $J_{0}(\nu r)\vect{e}_{z}$ leads to a simpler choice: 
$\Psi=(2\pi F_{0}/\nu^{3})\delta(\kappa_{z})(e^{i\nu p}+e^{-i\nu p})$, 
$\vect{\omega}=\vect{e}_{z}$. 

   We can easily show that $\vect{G}(p, \vect{\kappa})$, 
(\ref{Potentialequation}) is an eigenvector 

\begin{eqnarray} 
RBS[\vect{G}(q, \vect{\kappa})]
(p, \vect{\kappa})=-\vect{\kappa}\bm{\times}
[\vect{\kappa}\bm{\times}\vect{H}
(p, \vect{\kappa})]=\frac{1}{\nu}
\vect{G}(p, \vect{\kappa}),
\end{eqnarray}

\noindent of the $RBS$ operator (\ref{RadonBiotSavart}), 
if $\vect{\omega}$ is a fixed vector: 
$\partial_{p}\vect{\omega}=0$. 
Here $\vect{H}(p, \vect{\kappa})
=\Psi\vect{\omega}+(1/\nu)\vect{\Gamma}\bm{\times}
(\Psi\vect{\omega})$ is a potential for 
$\vect{G}(p, \vect{\kappa})$:
$\vect{G}(p, \vect{\kappa})
=\vect{\Gamma}\bm{\times}\vect{H}
(p, \vect{\kappa})$ and we use (\ref{oscillator}).

  The CK method also enables us to write a simple integral expression 
for the solution of (\ref{Radonfieldequation}). We employ in 
(\ref{Potentialequation}) integral expressions for simple harmonic 
oscillations (\ref{oscillator}). If we assume 
$\Psi_{k}(p, \vect{\kappa})$, ($k=1, \, 2$) are separable 
in $p$ and $\vect{\kappa}$, these are given as

\begin{eqnarray} \label{oscillatorrepresentation}
\Psi_{k}(p, \vect{\kappa})
=\frac{1}{4\pi i\nu} U_{k}(\vect{\kappa}) 
\int_{C^{\mp}_{\lambda}} \frac{e^{p\zeta}}{\zeta \mp i\lambda\nu} d\zeta,
\end{eqnarray}

\noindent on the infinite interval: $-\infty<p<\infty$, using
Laplace kernel: $e^{p\zeta}$ in the complex $\zeta$-plane.\cite{Ic, Zl}
Here $C^{\mp}_{\lambda}$ are arbitrary loops around the poles 
$\zeta=\pm i\lambda\nu$. This is basically the Cauchy integral 
representation formula for $e^{\pm i\lambda\nu p}$. We find 

\begin{eqnarray} \label{CKintegral}
\vect{G}(p, \vect{\kappa})
=\frac{1}{4\pi i} \left[ (i\lambda \vect{\kappa}\bm{\times}
\vect{\omega}_{1}
-\vect{\kappa} \bm{\times} \vect{\kappa}\bm{\times} 
\vect{\omega}_{1} )
\int_{C^{-}_{\lambda}} \frac{e^{p\zeta}}{\zeta - i\lambda\nu} d\zeta 
+(-i\lambda \vect{\kappa} \bm{\times} 
\vect{\omega}_{2}
-\vect{\kappa} \bm{\times} \vect{\kappa} \bm{\times} 
\vect{\omega}_{2} )
\int_{C^{+}_{\lambda}} \frac{e^{p\zeta}}{\zeta + i\lambda\nu} d\zeta \right],
\end{eqnarray}

\noindent using (\ref{oscillatorrepresentation}) 
in (\ref{Potentialequation}) and rearranging it. 
This is also oscillatory in $p$. We have included 
the $\vect{\kappa}$ dependent factors 
$U_{k}(\vect{\kappa})$ in the arbitrary 
functions $\vect{\omega}_{k}(\vect{\kappa})$. 
These are decoupled from the integral expressions. 
 
  For example, if we choose $\vect{\omega}_{1}
(\vect{\kappa})=[(2\pi)^{1/2}/(g\nu^{2})]
\vect{Q}_{\lambda}(\vect{\kappa})
s_{\lambda}(\lambda\nu \vect{\kappa})$ and 
$\vect{\omega}_{2}(\vect{\kappa})
=[(2\pi)^{1/2}/(g\nu^{2})]\vect{Q}_{\lambda}
(-\vect{\kappa})s_{\lambda}
(-\lambda\nu\vect{\kappa})$ 
in accordance with the Moses frame, we find
$\vect{G}(p, \vect{\kappa})
=\vect{F}^{\mathcal{R}}(p, \vect{\kappa})$,
(\ref{Radonform}).

  We can reproduce the L field (\ref{RadonLundquist}) choosing 
$\vect{\omega}_{1}\cong\delta(\kappa_{z})\vect{e}_{z}$ and 
$\vect{\omega}_{2}\cong\delta(\kappa_{z})\vect{e}_{z}$. We can 
also use $\vect{\omega}_{1}\cong\delta(\kappa_{z})\vect{e}_{\psi}$ 
and $\vect{\omega}_{2}\cong\delta(\kappa_{z})\vect{e}_{\alpha}$, 
($\kappa_{z}=\cos\alpha$, $\vect{\kappa}=\vect{e}_{r}$) on $\mathbb{S}^{2}$.

  If we substitute (\ref{CKintegral}) in the inversion formula 
(\ref{inverseRadon}), we find 

\begin{eqnarray} \label{integralphysical}
& & \vect{F}(\vect{x})
=\frac{1}{32\pi^{3}i}\nu^{2} 
\left[ \int_{C^{-}_{\lambda}} \frac{1}{\zeta - i\lambda\nu}
\int_{S^{2}} (i\lambda \vect{\kappa} \bm{\times} 
\vect{\omega}_{1}-\vect{\kappa} \bm{\times} 
\vect{\kappa} \bm{\times} \vect{\omega}_{1} )
e^{\zeta\vect{\kappa} \cdot \vect{x}} d\Omega\, 
d\zeta \right. \\ 
& & \hspace*{69mm} +\left.
\int_{C^{+}_{\lambda}} \frac{1}{\zeta + i\lambda\nu} 
\int_{S^{2}} (-i\lambda \vect{\kappa} \bm{\times} 
\vect{\omega}_{2}
-\vect{\kappa} \bm{\times} \vect{\kappa} \bm{\times} 
\vect{\omega}_{2} )
e^{\zeta\vect{\kappa} \cdot \vect{x}} d\Omega\, d\zeta
\right], \nonumber 
\end{eqnarray}

\noindent using the operator $\Gamma^{2}$ which is intertwined with 
$\nabla^{2}$, (\ref{transformvectorHelmholtz}) and interchanging the 
order of integrations. 

  If we choose

\begin{eqnarray}
& & \vect{\omega}_{1}=-i(2\pi)^{2}\frac{1}{\nu^{2}}
\left[ a\delta(\vect{\kappa}-\vect{\kappa}_{1})
\vect{E}_{1}
+b\delta(\vect{\kappa}-\vect{\kappa}_{2})
\vect{E}_{2} 
+c\delta(\vect{\kappa}-\vect{\kappa}_{3})
\vect{E}_{3} \right], \\
& & \vect{\omega}_{2}=-i(2\pi)^{2}\frac{1}{\nu^{2}}
\left[ a\delta(\vect{\kappa}+\vect{\kappa}_{1})
\vect{E}_{1}
+b\delta(\vect{\kappa}+\vect{\kappa}_{2})
\vect{E}_{2} 
+c\delta(\vect{\kappa}+\vect{\kappa}_{3})
\vect{E}_{3} \right], \nonumber
\end{eqnarray}

\noindent where $a$, $b$, $c$ are arbitrary constants, $i\lambda
\vect{\kappa}_{i}\bm{\times}\vect{E}_{i}
=\vect{E}_{i}$ and 
$\vect{\kappa}_{i}\cdot\vect{E}_{i}=0$, the equation
(\ref{integralphysical}) yields

\begin{eqnarray}
\vect{F}(\vect{x})
=-i \left( a e^{i\lambda\nu
\vect{\kappa}_{1}\cdot\vect{x}} 
\vect{E}_{1} 
+ b e^{i\lambda\nu
\vect{\kappa}_{2}\cdot\vect{x}} 
\vect{E}_{2} 
+ c e^{i\lambda\nu
\vect{\kappa}_{3}\cdot\vect{x}} 
\vect{E}_{3} \right).
\end{eqnarray}

\noindent The real part of this reduces to \textit{abc} 
field\cite{KS1} if we choose: $k_{1}=k_{2}=k_{3}=\lambda\nu$,
$\vect{\kappa}_{1}=\vect{e}_{z}$, $\vect{\kappa}_{2}=\vect{e}_{x}$,
$\vect{\kappa}_{3}=\vect{e}_{y}$ and 
$\vect{E}_{1}=\vect{e}_{x}+i\lambda\vect{e}_{y}$,
$\vect{E}_{2}=\vect{e}_{y}+i\lambda\vect{e}_{z}$,
$\vect{E}_{3}=\vect{e}_{z}+i\lambda\vect{e}_{x}$.
       
  We respectively refer the reader to Refs. \onlinecite{WM, RN} for Fourier 
integral expression on $\mathbb{S}^{2}$ of the general solution to the scalar 
Helmholtz equation and to Refs. \onlinecite{VK, VKVS, AL} for integral 
representations of the Beltrami fields in physical space.

\section{Topologically massive abelian gauge field}
\label{tmgf}

  The Euclidean topologically massive Abelian gauge 
field\cite{DJTS1, DJTS2, DJTS3} is a Trkalian field.\cite{KS1} 
Another example of Trkalian fields is the force-free magnetic 
field.\cite{GM} The Trkalian fields yield solutions of both the 
topologically massive gauge theories\cite{KS2, KS3, KS1} and
gravity.\cite{ANS} The spherical curl and Radon  transform analysis 
above are also valid for the Euclidean topologically massive Abelian 
gauge fields on $\mathbb{R}^{3}$. The gauge potential of the Trkalian 
field is also Trkalian. Meanwhile a gauge transformation corresponds 
to a vector normal to the sphere. The Radon transform of an 
\textit{anti-self-dual} potential (or field) is related by antipodal 
map on the sphere to the transform of the \textit{self-dual} potential 
obtained by inverting space coordinates.

  The Trkalian fields L and CK are also solutions of this theory. 
Furthermore, the L solution provides a simple example for quantization 
of the topological mass in this context.

\subsection{The gauge potential}

  We can easily derive the spherical curl and Radon transforms of 
the potential following the same\cite{Ml3} reasoning above. We find 

\begin{eqnarray} \label{simplifiedgaugepotential}
\vect{A}_{\lambda}(\vect{x})
=\frac{1}{g} \, \frac{1}{\nu} \, \int \vect{\chi}_{\lambda}
(\vect{x}|\lambda\nu \vect{\kappa})
s_{\lambda}(\lambda\nu \vect{\kappa}) d\Omega.
\end{eqnarray}

\noindent This satisfies the self-duality relation:
$\vect{F}_{\lambda}=\nu \vect{A}_{\lambda}$ where
$\vect{F}_{\lambda}=\vect{\nabla}\bm{\times} \vect{A}_{\lambda}$,
(\ref{simplifiedexpansion}). We also find

\begin{eqnarray} \label{Radonformpotential}
\vect{A}^{\mathcal{R}}_{\lambda}(p, \vect{\kappa})
=(2\pi)^{1/2} \, \frac{1}{g} \, \frac{1}{\nu^{3}} 
\left[ e^{i\lambda\nu p} \vect{Q}_{\lambda}(\vect{\kappa})
s_{\lambda}(\lambda\nu \vect{\kappa}) 
+ e^{-i\lambda\nu p} \vect{Q}_{\lambda}(-\vect{\kappa})
s_{\lambda}(-\lambda\nu \vect{\kappa}) \right].
\end{eqnarray}

\noindent These only differ by a factor\cite{KS1} 
of $1/\nu$ from those expressions for $\vect{F}_{\lambda}$ . 
Hence the potential $\vect{A}^{\mathcal{R}}_{\lambda}(p, \vect{\kappa})$, 
(\ref{Radonformpotential}) 
satisfies 

\begin{eqnarray} \label{equationtransform}
\vect{F}^{\mathcal{R}}_{\lambda}(p, \vect{\kappa})
-\nu\vect{A}^{\mathcal{R}}_{\lambda}(p, \vect{\kappa})=0,
\end{eqnarray}

\noindent where $\vect{F}^{\mathcal{R}}_{\lambda}
=\vect{\Gamma}\bm{\times}\vect{A}^{\mathcal{R}}_{\lambda}$, 
(\ref{Radonform}).

  A gauge transformation 

\begin{eqnarray} \label{gaugetransformation}
\vect{A}^{\prime}=\vect{A}-\frac{1}{g}\vect{\nabla}U,
\end{eqnarray}

\noindent of the potential\cite{KS1} 
is given by a curl-free vector

\begin{eqnarray} \label{gauge}
\vect{\nabla} U(\vect{x})=\frac{1}{\nu}
\int \vect{\chi}_{0}(\vect{x}|\vect{k})
f_{0}(\vect{k}) d^{3}k.
\end{eqnarray}

\noindent This yields the gauge function

\begin{eqnarray}
U(\vect{x})
=\frac{1}{(2\pi)^{3/2}} \,\, i \, \frac{1}{\nu}
\int e^{i \vect{k} \cdot \vect{x}} 
f_{0}(\vect{k}) \frac{1}{k} d^{3}k. 
\end{eqnarray}

\subsection{The anti-self-dual case}
\label{Antiselfdual}

  The \textit{anti-self-dual} case 

\begin{eqnarray} \label{asdfieldequation}
\vect{\nabla}\bm{\times}\vect{F}+\nu\vect{F}=0,
\end{eqnarray}

\noindent of equation (\ref{fieldequation}) contains an extra 
factor of ($-$) in $\nu$. This terminology is motivated by the 
interchange symmetry $\vect{F}\leftrightarrow\nu\vect{A}$.
This leads to a flip of sign in the spherical curl and Radon transforms 
of the self-dual field (\ref{fieldequation}). In this case, an arbitrary 
solution is given in terms of its transform on the sphere of radius 
$k=-\lambda\nu=|\nu|$. Furthermore, only the eigenfunctions for which 
$\lambda=-sgn(\nu)$ contribute to the field. We find 

\begin{eqnarray} \label{asdsimplifiedexpansion}
& & \vect{F}_{\lambda}(\vect{x})
=\frac{1}{g} \int \vect{\chi}_{\lambda}
(\vect{x}|-\lambda\nu \vect{\kappa})
s_{\lambda}(-\lambda\nu \vect{\kappa}) d\Omega \\
& & \hspace*{11mm}
=\frac{1}{(2\pi)^{3/2}} \,\,  \frac{1}{g}
\int e^{-i \lambda \nu \vect{\kappa} \cdot \vect{x}} 
\vect{Q}_{\lambda} (\vect{\kappa})
s_{\lambda}(-\lambda\nu\vect{\kappa}) d\Omega, \nonumber
\end{eqnarray}

\noindent and

\begin{eqnarray} \label{asdsphericalcurltransform}
s_{\lambda}(-\lambda\nu\vect{\kappa})
=\frac{1}{(2\pi)^{1/2}} g\nu^{2} e^{i\lambda\nu p} 
F^{\mathcal{R}}_{a}(p, \vect{\kappa}).
\end{eqnarray}

\noindent The Radon transform 

\begin{eqnarray} \label{asdRadonform}
\vect{F}^{\mathcal{R}}_{\lambda}(p, \vect{\kappa})
=(2\pi)^{1/2} \, \frac{1}{g} \, \frac{1}{\nu^{2}} 
\left[ e^{i\lambda\nu p} \vect{Q}_{\lambda}(-\vect{\kappa})
s_{\lambda}(\lambda\nu \vect{\kappa}) 
+ e^{-i\lambda\nu p} \vect{Q}_{\lambda}(\vect{\kappa})
s_{\lambda}(-\lambda\nu \vect{\kappa}) \right],
\end{eqnarray}

\noindent of the field satisfies: $\vect{\Gamma}\bm{\times}
\vect{F}^{\mathcal{R}}_{\lambda}=-\nu\vect{F}^{\mathcal{R}}_{\lambda}$.
Noting $\lambda\nu=|\nu|>0$ for the self-dual case and $-\lambda\nu=|\nu|>0$ 
for the anti-self-dual case, the equation (\ref{asdRadonform}) coincides 
with (\ref{Radonform}). The difference of the two cases is the relative 
sign of $p$ and $\vect{\kappa}$. We shall present a comparison of the 
two cases below.

  The gauge potential

\begin{eqnarray} \label{asdsimplifiedgaugepotential}
\vect{A}_{\lambda}(\vect{x})
=-\frac{1}{g} \, \frac{1}{\nu} \, \int \vect{\chi}_{\lambda}
(\vect{x}|-\lambda\nu \vect{\kappa})
s_{\lambda}(-\lambda\nu \vect{\kappa}) d\Omega, 
\end{eqnarray}

\noindent for the field (\ref{asdsimplifiedexpansion}) 
satisfies the anti-self-duality relation: 
$\vect{F}_{\lambda}=-\nu\, \vect{A}_{\lambda}$.
Its Radon transform 
 
\begin{eqnarray} \label{asdRadonformasdpotential}
\vect{A}^{\mathcal{R}}_{\lambda}(p,\vect{\kappa})
=-(2\pi)^{1/2} \, \frac{1}{g} \, \frac{1}{\nu^{3}} 
\left[ e^{i\lambda\nu p} \vect{Q}_{\lambda}(-\vect{\kappa})
s_{\lambda}(\lambda\nu \vect{\kappa}) 
+ e^{-i\lambda\nu p} \vect{Q}_{\lambda}(\vect{\kappa})
s_{\lambda}(-\lambda\nu \vect{\kappa}) \right],
\end{eqnarray}

\noindent satisfies: $\vect{F}^{\mathcal{R}}_{\lambda}
=-\nu\vect{A}^{\mathcal{R}}_{\lambda}$.

\subsection{The Radon transform and duality classes}
\label{asdtrf}

  We can compare the Radon transforms of self-dual and anti-self-dual 
fields using a correspondence between them. A self-dual field becomes 
anti-self-dual, or vice versa under inversion of space coordinates, 
if the topological mass is held fixed.\cite{Br} (Hence the topological 
mass is actually a pseudoscalar.) Therefore, given a self-dual-field 
$\vect{F}(\vect{x})$:
$\vect{\nabla}_{x}\bm{\times}\vect{F}
(\vect{x})=\nu\,\,\vect{F}
(\vect{x})$, we can write the corresponding anti-self-dual 
field as $\vect{F}^{\prime}(\vect{x})
=\vect{F}(-\vect{x})$:
$\vect{\nabla}_{x}\bm{\times}
\vect{F}^{\prime}(\vect{x})
=-\nu\,\,\vect{F}^{\prime}(\vect{x})$
inverting the coordinates:
$\vect{x} \longrightarrow \vect{x}^{\prime}
=- \vect{x}$. 

  If $T$ is (matrix of) a nonsingular linear transformation: 
$T^{-1}\vect{x}=\vect{x}^{\prime}$,\cite{GGV} 
then 

\begin{eqnarray} \label{Radonvectorinversion}
& & \vect{F}^{\prime\,\vect{\mathcal{R}}}
(p, \vect{\kappa})
=\vect{\mathcal{R}}[\vect{F}^{\prime}(\vect{x})]
(p, \vect{\kappa})
=\vect{\mathcal{R}}[\vect{F}(T^{-1}\vect{x})]
(p, \vect{\kappa}) \\
& & \hspace*{17mm} =\int
\vect{F}(T^{-1}\vect{x})
\delta(p-\vect{\kappa}\cdot\vect{x})
d^{3}x \nonumber \\
& & \hspace*{17mm} =|T|\int
\vect{F}(\vect{x}^{\prime})
\delta(p-T^{\dagger}\vect{\kappa}\cdot\vect{x}^{\prime})
d^{3}x^{\prime} \nonumber \\
& & \hspace*{17mm} =|T|\vect{F}^{\mathcal{R}}
(p, T^{\dagger}\vect{\kappa}).
\nonumber
\end{eqnarray}

\noindent The inversion of coordinates is given by $T=-1$, $|T|=|det[T]|=1$ 
which is an orthogonal matrix: $TT^{\dagger}=1$. Thus, the Radon transform of 
an anti-self-dual field 

\begin{eqnarray} \label{Radonantiselfdual}
\vect{F}^{\prime\,\vect{\mathcal{R}}}
(p, \vect{\kappa})=
\vect{F}^{\mathcal{R}}
(p, -\vect{\kappa}),
\end{eqnarray}

\noindent is related by antipodal map on the sphere 
to the transform of the self-dual field obtained by 
inverting the coordinates.

  In the Moses coframe, we can find the Radon transform of an 
anti-self-dual field similarly inverting the coordinates in 
(\ref{Radoncomp}). This leads to the 
inversion $\vect{\kappa}\longrightarrow -\vect{\kappa}$ 
in (\ref{Radonform}). A flip of sign (\ref{identities}): 
$p\longrightarrow -p$ or $\vect{\kappa}
\longrightarrow -\vect{\kappa}$ (but not both) brings 
$*\,{^{a}}{F}^{R}(p, -\vect{\kappa})$ back 
to (\ref{Radonform}) with $\lambda\nu=|\nu|>0$ or (\ref{asdRadonform}) 
with $-\lambda\nu=|\nu|>0$. Note that the equation (\ref{Radonantiselfdual}) 
together with (\ref{Radonform}) yields a representation different from 
(\ref{asdRadonform}) for anti-self-dual fields. These coincide 
upon both: $\nu\longrightarrow-\nu$ and 
$\vect{\kappa}\longrightarrow-\vect{\kappa}$.

   The Radon transform of an anti-self-dual field or potential 
satisfy the equations (\ref{transelfdualfield}) or (\ref{equationtransform})
with an opposite sign for the topological mass as noted above. A simple 
example is given by $\vect{F}(\vect{x})=e^{i\mu\lambda\nu\vect{\kappa}_{0}\cdot
\vect{x}} \vect{Q}_{\lambda}(\vect{\kappa}_{0})$ where $\mu=\pm 1$ correspond 
to the self-dual and anti-self-dual cases. We find

\begin{eqnarray} \label{exampleoppositehemispheresecond}
\vect{F}^{\mathcal{R}}(p, \vect{\kappa})
=(2\pi)^{2} \, \frac{1}{\nu^{2}} 
\left[ e^{i\lambda\nu p} 
\delta(\vect{\kappa}-\mu\vect{\kappa}_{0})
+e^{-i\lambda\nu p} 
\delta(\vect{\kappa}+\mu\vect{\kappa}_{0}) \right]
\vect{Q}_{\lambda}(\vect{\kappa}_{0}),
\end{eqnarray}

\noindent using $\lambda\nu>0$ in the self-dual case 
(\ref{exampleoppositehemispherefirst}) and $-\lambda\nu>0$ 
in the anti-self-dual case. This yields
$\vect{\Gamma}\bm{\times} \vect{F}^{\mathcal{R}}
=\mu\nu\vect{F}^{\mathcal{R}}$ anticipating the 
effect of $\delta(\vect{\kappa}\mp\mu\vect{\kappa}_{0})$.\cite{Ml3}

  Noting the inversion of the refined transformation for a self-dual-field 
$\vect{F}^{R}(p, \vect{\kappa})$ on a canonical hemisphere in Subsection 
\ref{Ref}, one can require $\vect{\kappa}$ to be restricted into 
the complementary hemisphere for the anti-self-dual field 
$\vect{F}^{\prime\,{\mathcal{R}}}(p, \vect{\kappa})$. 
(Also see Appendix \ref{RadontrfRadon}.)

\subsection{The Lundquist Solution}

  The constant eigenvalue $\nu$, called the topological mass in 
this context, is intimately related to boundary conditions imposed 
on the field\cite{LS, C, CK} and the global topology determined by 
them. The quantization of the topological mass here arises as a result 
of a well defined gauge transformation as in the previous\cite{KS2, KS1} 
examples. 

  The eigenvalue in L solution determines the twist per unit length of 
the field lines on the $z$-axis, see for example Refs. \onlinecite{BYH, DMDL}. 
In the topologically massive case, the strength of the field on the $z$-axis 
is proportional to the square of the eigenvalue. Meanwhile the gauge potential 
is of strength proportional to this twisting. This analogy suggests a natural 
geometric interpretation for the Lundquist field in this context. 

  We find the potential

\begin{eqnarray}
\vect{A}=\frac{1}{\nu}\vect{F}_{L}
-\frac{1}{\nu}F_{0}\,\vect{e}_{z},
\end{eqnarray}

\noindent integrating\cite{GP} the components of $\vect{F}_{L}$, 
(\ref{lundquist}). This satisfies: $\vect{\nabla}\bm{\times}\vect{A}-\nu 
\vect{A}=F_{0}\vect{e}_{z}$ with an extra term.  

  In the topologically massive Abelian gauge theory,\cite{KS2, KS1} 
we can make this term vanish: $\vect{F}^{\prime}_{L}
-\nu\vect{A}^{\prime}=0$ by a gauge transformation:
$\vect{A}^{\prime}=\vect{A}-(i/g)\vect{\nabla}\ln{U}$
identifying it with a gauge term: 
$-(i/g)\vect{\nabla}\ln{U}=(1/\nu)F_{0}\vect{e}^{3}$. This yields 
$U=e^{i\nu z}$ which takes values in the group $U(1)$, 
choosing $F_{0}=\nu^{2}/g$. The strength of the gauge potential is 
given by the gauge coupling constant $\nu/g=ng$, if $\nu=ng^{2}$. 
This leads us to adopt a fundamental scale\cite{KS2, KS1} of 
length $l=2\pi/g^{2}$. We can write $\nu=ng^{2}$ as $\nu=2\pi n/l$. 
If $U=e^{i(2\pi n/l)z}$ is a single-valued function of $z$ with 
the fundamental scale $l$, then $n$ has to be an integer. The fundamental 
length scale $l$ is the least common multiple of intervals over which the 
gauge function is single-valued and periodic for any integer $n$, in addition 
to the fact that it has a smaller period $l^{\prime}=l/n$.\cite{KS2, KS1} 
The discussion of physical and topological aspects of this solution and 
the effect of gauge transformations on these would be distracting us from 
our purposes here.

  The anti-self-dual solution is given by $\vect{\kappa}
\longrightarrow -\vect{\kappa}$ in (\ref{RadonLundquist}). 
Also note that the L field defines a contact structure 
since $J_{0}$ and $J_{1}$ have no common zeros.\cite{AGGM} 

\section{CONCLUSION}

  The spherical curl transformation for Trkalian fields 
is a Radon probe transformation in the Moses frame. 
We have written this using differential forms. This is 
an example of an integral transform which can be formally 
written this way.

  The point of view here suggests an approach for studying the Trkalian 
fields in the transform space. The transform space representation has 
certain advantages. First of all, the equation in the transform 
space provides a concrete frame for studying these fields, for example 
using the spherical curl transform which is based on Moses eigenfunctions. 
This further offers a simplification since the differential and algebraic 
aspects are separable in the transform space. This also exhibits the role 
of the Moses eigenbasis which yields a helicity decomposition of the Radon 
transform. Meanwhile the field can be reconstructed using knowledge of the 
transform on a canonical hemisphere. Briefly, this equation provides 
a simple geometric frame in the transform space for Trkalian fields 
with a new insight into their structure besides practical advantages.  

  We have discussed the connection of the Radon transformation 
with the $BS$ integral. We have introduced the $RBS$ operator in the 
transform space. This is simply given in terms of $1$-dimensional Fourier 
transforms. We can also associate an eigenvalue equation for $RBS$ 
operator with the Trkalian fields in transform space. The Radon transform 
of a Trkalian field is an eigenvector of this operator. Its kernel consists 
of those vector fields which are normal to the sphere. We have also presented 
an Ampere law type relation for these fields.

  We have presented applications of these on the L solution. Then 
we have discussed the CK method with circular and elliptic solutions. 
The elliptic one reduces to VR solution. The L solution also defines 
a contact structure.

   We have presented CK solution of the corresponding eigenvalue equation 
in the transform space. The CK method in transform space also leads to 
a simplification eliminating the differential operations and hence 
reducing the solution to simple algebraic manipulations consisting of 
arbitrary but fixed vectors. Because the Radon transform reduces the 
scalar Helmholtz equation to that of simple harmonic oscillations. The 
poloidal, toroidal representations also suggest the use of the Moses 
eigenbasis. This simplification also enables us to write integral 
representations as we have seen on a simple example. One can try 
imposing different boundary conditions and symmetries motivated 
by considerations in the physical space. The CK solution is 
also an eigenvector of the $RBS$ operator. 

  The Euclidean topologically massive Abelian gauge fields and also the 
force-free magnetic fields are examples of Trkalian fields. Hence the 
Radon and the spherical curl transform analysis are also valid for these. 
In the topologically massive case, the gauge potential is also Trkalian. 
The Radon transform of an \textit{anti-self-dual} potential (or field) 
is related by antipodal map on the sphere to the transform of the 
\textit{self-dual} potential obtained by inverting space coordinates.

  The L and CK fields are also solutions of this theory. This has 
been overlooked previously. Furthermore, the L solution provides 
an example for quantization of the topological mass in this theory. 
This suggests a natural geometric interpretation for the Lundquist 
field in this context. However the discussion of physical implications 
of this is beyond our goals here.

\begin{acknowledgments}

  The author would like to thank the anonymous referee 
and the Editor for clear comments and guidance.

\end{acknowledgments}

\appendix

\section{THE RADON TRANSFORMATION}
\label{Radontrf}
 
\subsection{The Radon transformation}
\label{RadontrfRadon}

  The Radon transform of vector fields is defined\cite{Ml3} as

\begin{eqnarray} \label{Radonvectortransform}
\vect{F}^{\mathcal{R}}(p, \vect{\kappa})
=\vect{\mathcal{R}}[\vect{F}(\vect{x})]
(p, \vect{\kappa})
=\int\vect{F}(\vect{x})
\delta(p-\vect{\kappa}\cdot\vect{x})d^{3}x.
\end{eqnarray}

\noindent This is a componentwise generalization 
of Radon transform 

\begin{eqnarray}
f^{\mathcal{R}}(p, \vect{\kappa})
=\vect{\mathcal{R}}[f(\vect{x})](p, \vect{\kappa})
\int f(\vect{x})
\delta(p-\vect{\kappa}\cdot\vect{x})d^{3}x,
\end{eqnarray}

\noindent of scalar functions in cartesian coordinates. 
This satisfies the identities 

\begin{eqnarray}
f^{\mathcal{R}}(-p, -\vect{\kappa})=f^{\mathcal{R}}(p, \vect{\kappa}), 
\hspace*{8mm} 
f^{\mathcal{R}}(-p, \vect{\kappa})=f^{\mathcal{R}}(p, -\vect{\kappa}).
\end{eqnarray}

\noindent The Radon transform (\ref{Radonvectortransform}) of vectors also 
satisfies similar identities.

  The Radon transform is simply defined\cite{GGV, Helgason} for (vector) 
functions in Schwartz class $\mathbf{{\mathcal{S}}}[\mathbb{R}^{3}]$ of 
rapidly decreasing functions on $\mathbb{R}^{3}$. Hence the function 
vanishes at infinity as we assume repeatedly, for example in Sections 
\ref{Radontrfs} and \ref{RBS}. Then the Radon transform is a function 
in $\mathbf{{\mathcal{S}}}[\mathbb{S}^{2}\bm{\times}\mathbb{R}]$. A natural 
domain of definition for the Radon transform is the set of hyperplanes in 
$\mathbb{R}^{3}$, i. e. the projective space. The space 
$\mathbb{S}^{2}\bm{\times}\mathbb{R}$ is a two-fold 
covering of this.\cite{Helgason, Markoe, RK} Note that the definition 
(\ref{Radonvectortransform}) actually provides an extension\cite{Markoe, N} 
of the Radon transform as a function homogeneous of degree $-1$ on 
$(\mathbb{R}^{3}-\{0\})\bm{\times}\mathbb{R}$. The adjoint transform, 
see Definition \ref{adjointRadon}, works in the opposite direction. 
We refer the reader to Refs. \onlinecite{P1, Helgason, Markoe, N} 
or \onlinecite{D} for a derivation of (\ref{adjointRadonRadon}).
The inverse Radon transform is given as

\begin{eqnarray} \label{inverseRadon}
\vect{F}(\vect{x})=-\frac{1}{8\pi^{2}}
\nabla^{2}_{x} \int_{S^{2}} 
\vect{F}^{\mathcal{R}}(\vect{\kappa}\cdot\vect{x}, 
\vect{\kappa}) d\Omega,
\end{eqnarray}

\noindent where $\nabla^{2}_{x}$ is the Laplace operator.

  We can prove the Fourier slice theorem (\ref{Fourierslice}) 
using (\ref{Radonvectortransform}). The relation (\ref{FourierRiesz}) 
for Fourier transform of the Riesz potential is valid for functions in 
the Schwartz class.\cite{Markoe}

  The Radon probe transformation\cite{NW} is defined as 

\begin{eqnarray}
{\mathfrak{R}}[\vect{F}(\vect{x})]
(p, \vect{\kappa})
=\vect{V}(p, \vect{\kappa})\cdot
\vect{\mathcal{R}}[\vect{F}(\vect{x})]
(p, \vect{\kappa}),
\end{eqnarray}

\noindent where $\vect{V}(p, \vect{\kappa})$ is the probe. 

  Further note that, if we had defined the Radon transformation 
on the projective space of hyperplanes\cite{Helgason} in $\mathbb{R}^{3}$, 
then in Section \ref{tmgf} we would take the Radon transforms of self-dual 
and anti-self-dual fields on distinct spaces.

\subsection{Inversion of the refined transformation}
\label{RadontrfRefinement}

  The Radon transform (\ref{Radonvectortransform}) is a right 
inverse: $\vect{\mathcal{R}}^{-1}_{H}\vect{\mathcal{R}}[\vect{F}]
(\vect{x})=\vect{F}(\vect{x})$ of the refined\cite{MP} 
(inverse) transformation 

\begin{eqnarray} 
& & \vect{\mathcal{R}}^{-1}_{H}[\vect{F}^{\mathcal{R}}
(p, \vect{\kappa})](\vect{x}) 
=-\frac{1}{4\pi^{2}} \nabla^{2}
\int_{H} \vect{F}^{\mathcal{R}}
(\vect{\kappa}\cdot\vect{x}, \vect{\kappa}) 
d\Omega \\
& & \hspace*{29mm} 
=-\frac{1}{4\pi^{2}} \nabla^{2}
\int \int_{H} \vect{F}^{\mathcal{R}}
(p, \vect{\kappa}) \delta(p-\vect{\kappa}\cdot\vect{x})
d\Omega dp, \nonumber 
\end{eqnarray}

\noindent for any canonical hemisphere $H$.
However as a left inverse, we find
 
\begin{eqnarray}
& & \vect{\mathcal{R}}\vect{\mathcal{R}}^{-1}_{H}[\vect{F}
(p^{\prime}, \vect{\kappa}^{\prime})](p, \vect{\kappa})
=-\int \int_{H} \vect{F}(p^{\prime}, \vect{\kappa}^{\prime})
W(p, \vect{\kappa}; p^{\prime}, \vect{\kappa}^{\prime}) d\Omega^{\prime} 
dp^{\prime} \\
& & \hspace*{35mm}
=\left\{ \begin{array}{l}
\vect{F}(p, \vect{\kappa}), \,\,\,\, \vect{\kappa}\in H, \\
\vect{F}(-p, -\vect{\kappa}), \,\,\,\, \vect{\kappa}\in H^{\prime},
\end{array} \right. \nonumber
\end{eqnarray}

\noindent where 

\begin{eqnarray}
& & W(p, \vect{\kappa}; p^{\prime}, \vect{\kappa}^{\prime})
=\frac{1}{(2\pi)^{2}} \int \delta(p-\vect{\kappa}\cdot\vect{x})
\nabla^{2}_{x} \delta(p^{\prime}-\vect{\kappa}^{\prime}\cdot\vect{x})
d^{3}x \\
& & \hspace*{23mm} 
=-\delta(p^{\prime}-p)\delta(\vect{\kappa}^{\prime}-\vect{\kappa})
-\delta(p^{\prime}+p)\delta(\vect{\kappa}^{\prime}+\vect{\kappa}). 
\nonumber
\end{eqnarray}

\noindent  We refer the reader to Ref. \onlinecite{MP} for evaluation 
of this integral. Thus only in the case $\vect{F}(p, \vect{\kappa})$ 
is the Radon transform of a function can we define both right 
and left inverse of this refined transformation.\cite{MP} 

\section{THE SPHERICAL CURL AND RADON TRANSFORMS}
\label{SphCurlRadon}

  The inverse Radon transform in differential forms is given as

\begin{eqnarray} \label{inverseRadondif}
*\tensor*[^{a}]{F}{}(\vect{x})
=\frac{1}{8\pi^{2}}
\triangle_{x} \int_{S^{2}} 
*\tensor*[^{a}]{F}{}^{\mathcal{R}}(\vect{\kappa} \cdot \vect{x}, 
\vect{\kappa}) d\Omega,
\end{eqnarray}

\noindent where $\triangle_{x}=(d+\delta)^{2}=\delta d +d \delta$ is the 
Laplace-Beltrami operator and $\delta \omega_{r}=(-1)^{r}*d*\omega_{r}$ 
for an $r$-form $\omega_{r}$. 

  We can prove equivalence of the spherical curl transform and 
the Radon transform as follows.\cite{Ml3} If we substitute 
(\ref{sphericalcurltransform}) in (\ref{simplifiedexpansion}), 
we find 

\begin{eqnarray} \label{SphericalcurlRAdon1}
& & *\tensor*[^{a}]{F}{}(\vect{x})
=\frac{1}{(2\pi)^{2}} \, \nu^{2} \,  
\int e^{-i \lambda \nu (p-\vect{\kappa} \cdot \vect{x})} 
\omega^{a} (\vect{\kappa})
F^{\mathcal{R}}_{a}(p, \vect{\kappa}) d\Omega.
\end{eqnarray}

\noindent We can write this as

\begin{eqnarray} 
*\tensor*[^{a}]{F}{}(\vect{x})
=\frac{1}{(2\pi)^{2}} \, \nu^{2}\, dx^{i}
\int e^{-i \lambda \nu (p-\vect{\kappa} \cdot \vect{x})} 
|\omega^{a}_{i} (\vect{\kappa})>
<\omega^{a}_{k}(\vect{\kappa}), 
*{^{a}}{\mathcal{F}}^{\mathcal{R}}_{k}(p, \vect{\kappa})>
d\Omega. \nonumber
\end{eqnarray}

\noindent We find 

\begin{eqnarray} \label{nodyadic}
\int e^{-i \lambda \nu (p-\vect{\kappa} \cdot \vect{x})} 
|\omega^{a}_{i} (\vect{\kappa})>
<\omega^{a}_{k}(\vect{\kappa}), 
*\tensor*[^{a}]{\mathcal{F}}{}^{\mathcal{R}}_{k}(p, \vect{\kappa})>
d\Omega =
\frac{1}{2}
\int e^{-i \lambda \nu (p-\vect{\kappa} \cdot \vect{x})}
\delta_{ik} | *\tensor*[^{a}]{\mathcal{F}}{}^{\mathcal{R}}_{k}
(p, \vect{\kappa})> d\Omega,
\end{eqnarray}

\noindent following a similar reasoning given in Ref. \onlinecite{Ml3} 
with our notation. Here we use the completeness relation (\ref{orthocompfin}) 
for the coframe, the expression (\ref{Radonform}) for the Radon transform of 
the field which yields 
$<\omega^{3}, *\tensor*[^{a}]{\mathcal{F}}{}^{\mathcal{R}}>=0$ 
and the fact that the integrals of the two remaining terms over 
the sphere are equal. Then we are led to 

\begin{eqnarray} \label{SphericalcurlRAdon2}
& & *\tensor*[^{a}]{F}{}(\vect{x})
=\frac{1}{2(2\pi)^{2}} \, \nu^{2} \,  
\int e^{-i \lambda \nu (p-\vect{\kappa} \cdot \vect{x})} 
*\tensor*[^{a}]{F}{}^{\mathcal{R}}(p, \vect{\kappa}) d\Omega.
\end{eqnarray}

\noindent One recovers the inverse Radon transform formula 
(\ref{inverseRadondif}) introducing the Laplace-Beltrami operator 
and restricting $p=\vect{\kappa}\cdot\vect{x}$.\cite{Ml3}

   Also substituting (\ref{Radonform}) for 
$p=\vect{\kappa}\cdot\vect{x}$ 
in (\ref{inverseRadondif})

\begin{eqnarray} \label{RadonSphercurl}
*\tensor*[^{a}]{F}{}(\vect{x})
=\frac{1}{2(2\pi)^{3/2}} \, \frac{1}{g} \, \frac{1}{\nu^{2}} 
\triangle_{x} \int \left[ 
e^{i\lambda\nu\vect{\kappa}\cdot\vect{x}} 
\omega^{a}(\vect{\kappa})
s_{a}(\lambda\nu \vect{\kappa}) 
+ e^{-i\lambda\nu\vect{\kappa}\cdot\vect{x}} 
\omega^{a}(-\vect{\kappa})
s_{a}(-\lambda\nu \vect{\kappa}) \right] d\Omega,
\end{eqnarray}

\noindent we see that the inverse Radon transform 
(\ref{inverseRadondif}) yields the expression 
(\ref{simplifiedexpansion}) of $*\tensor*[^{a}]{F}{}(\vect{x})$
in terms of the spherical curl transform.\cite{Ml3} 
Here the second integral reduces to the first one 
as $\vect{\kappa}\longrightarrow -\vect{\kappa}$. 

\section{BIOT-SAVART INTEGRAL OPERATOR}
\label{BS}

i)
\vspace*{-7mm}
\begin{eqnarray}
\hspace*{-129mm}
\vect{\nabla}_{x} \cdot 
BS[\vect{F}](\vect{x})=0,
\nonumber
\end{eqnarray}

\begin{eqnarray}
& & \vect{\nabla} \cdot 
BS[\vect{F}](\vect{x})
=\vect{\nabla} \cdot 
\vect{\nabla} \bm{\times}
A[\vect{F}](\vect{x}) \\
& & \hspace{23mm} =\frac{1}{4\pi} \vect{\nabla} \cdot 
\int_{D} \vect{F}(\vect{y})
\bm{\times} \frac{\vect{x}-\vect{y}}
{|\vect{x}-\vect{y}|^{3}} \, d^{3}y \nonumber \\
& & \hspace{23mm} =0. \nonumber 
\end{eqnarray}

\vspace*{2mm}

ii)
\vspace*{-7mm}
\begin{eqnarray}
\hspace*{-118mm}
\vect{\nabla}_{x} \bm{\times} 
BS[\vect{F}](\vect{x})
=\vect{F}(\vect{x}),
\nonumber
\end{eqnarray}

\begin{eqnarray}
& & \vect{\nabla}_{x} \bm{\times} 
BS[\vect{F}](\vect{x})
=\vect{\nabla}_{x} \bm{\times} \vect{\nabla}_{x} 
\bm{\times} A[\vect{F}](\vect{x}) \\
& & \hspace{27mm} =\frac{1}{4\pi} 
\vect{\nabla}_{x} \bm{\times} \vect{\nabla}_{x} \bm{\times} 
\int_{D} \frac{\vect{F}(\vect{y})}
{|\vect{x}-\vect{y}|} \, d^{3}y, \nonumber \\
& & \hspace{27mm} =-\frac{1}{4\pi} \nabla^{2}_{x} \int_{D} 
\frac{\vect{F}(\vect{y})}
{|\vect{x}-\vect{y}|} \, d^{3}y
+\frac{1}{4\pi} \vect{\nabla}_{x} \int_{D}
\vect{F}(\vect{y}) 
\cdot \vect{\nabla}_{x}
\frac{1}{|\vect{x}-\vect{y}|} \, d^{3}y 
\nonumber \\
& & \hspace{27mm} =-\frac{1}{4\pi} \int_{D} 
\vect{F}(\vect{y})
\nabla^{2} \frac{1}{|\vect{x}-\vect{y}|} \, d^{3}y 
-\frac{1}{4\pi} \vect{\nabla}_{x} \int_{D}
\vect{F}(\vect{y}) \cdot 
\vect{\nabla}_{y}
\frac{1}{|\vect{x}-\vect{y}|} \, d^{3}y.
\nonumber 
\end{eqnarray}

\noindent We can write the second integral as

\begin{eqnarray}
& & \int_{D} \vect{F}(\vect{y}) 
\cdot \vect{\nabla}_{y}
\frac{1}{|\vect{x}-\vect{y}|} \, d^{3}y
=\int_{D} \vect{\nabla}_{y} \cdot \left[
\frac{\vect{F}(\vect{y})}
{|\vect{x}-\vect{y}|} \right] \, d^{3}y
-\int_{D} \frac{\vect{\nabla}_{y}\cdot
\vect{F}(\vect{y})}
{|\vect{x}-\vect{y}|}\, d^{3}y  \\
& & \hspace{40mm} =-\int_{D} \frac{\vect{\nabla}_{y} 
\cdot \vect{F}(\vect{y})}
{|\vect{x}-\vect{y}|}\, d^{3}y
+ \oint_{S} \frac{\vect{F}(\vect{y})}
{|\vect{x}-\vect{y}|} \cdot d\vect{s}.
\nonumber
\end{eqnarray}

\noindent This yields

\begin{eqnarray}
& & \vect{\nabla}_{x} \bm{\times} 
BS[\vect{F}](\vect{x})
=\vect{F}(\vect{x})
+\frac{1}{4\pi} \vect{\nabla}_{x} 
\int_{D} \frac{\vect{\nabla}_{y}\cdot
\vect{F}(\vect{y})}
{|\vect{x}-\vect{y}|}\, d^{3}y 
-\frac{1}{4\pi} \vect{\nabla}_{x}
\oint_{S} \frac{\vect{F}(\vect{y})}
{|\vect{x}-\vect{y}|} \cdot d\vect{s}_{y} \\
& & \hspace{26mm} =\vect{F}(\vect{x})
-\frac{1}{4\pi} \int_{D} \vect{\nabla}_{y} \cdot 
\vect{F}(\vect{y})
\frac{\vect{x}-\vect{y}}
{|\vect{x}-\vect{y}|^{3}} \, d^{3}y 
+\frac{1}{4\pi} \oint_{S} \frac{\vect{x}-\vect{y}}
{|\vect{x}-\vect{y}|^{3}} 
\vect{F}(\vect{y})\cdot d\vect{s}_{y}, 
\nonumber
\end{eqnarray}

\noindent for $\vect{x}\in D$. Thus we conclude

\begin{eqnarray}
\vect{\nabla}_{x} \bm{\times} 
BS[\vect{F}](\vect{x})
=\vect{F}(\vect{x}), \hspace*{5mm}
\vect{x} \in D, 
\end{eqnarray}

\noindent if 

  a) $\vect{F}(\vect{y})$ 
is divergence-free: $\vect{\nabla}_{y} \cdot 
\vect{F}(\vect{y})=0$,

  b) $\vect{F}(\vect{y})$ 
vanishes at infinity, as the region $D$ is extended to 
whole space $\mathbb{R}^{3}$ [for a finite region $D$, 
$\vect{F}(\vect{y})$ 
is tangent \\ 
\hspace*{7mm} to the surface $S=\partial D$ bounding 
this region]. 

\section{BIOT-SAVART INTEGRAL OF THE LUNDQUIST FIELD}
\label{BSL}

  We shall evaluate 

\begin{eqnarray} \label{BiotSavartLundquist}
& & BS[\vect{F}_{L}](\vect{x})
=\frac{1}{4\pi}
\int_{D} \vect{F}_{L}(\vect{y})
\bm{\times} \frac{\vect{x}-\vect{y}}
{|\vect{x}-\vect{y}|^{3}} \, d^{3}y \\
& & \hspace*{19mm} = I_{1}+I_{2}+I_{3}+I_{4}+I_{5},
\nonumber
\end{eqnarray}

\noindent which can be decomposed into five terms. We choose: 
$\vect{y}=r\vect{e}_{r}(\phi)+z\vect{e}_{z}$
and $\vect{x}=R\vect{e}_{r}(\theta)$ since 
$\vect{F}_{L}$ is invariant along the 
$z$-axis. We shall need 

\begin{eqnarray} \label{Integral}
\int_{z=-\infty}^{\infty} \frac{z}{(a^{2}+z^{2})^{3/2}}dz=0,
\hspace*{10mm}
\int_{z=-\infty}^{\infty} \frac{1}{(a^{2}+z^{2})^{3/2}}dz=2\frac{1}{a^{2}}.
\end{eqnarray}

\noindent We shall also make use of Poisson integral formulas

\begin{eqnarray} \label{Poissonintegral}
& & \frac{1}{2\pi}\int_{\phi=0}^{2\pi} \, P(R, r, \phi-\theta)d\phi=1, 
\hspace*{5mm}
\frac{1}{2\pi}\int_{\phi=0}^{2\pi} R\sin\phi \, P(R, r, \phi-\theta)d\phi
=r\sin\theta, \\
& & \hspace*{52mm} \frac{1}{2\pi}\int_{\phi=0}^{2\pi} R\cos\phi \, 
P(R, r, \phi-\theta)d\phi
=r\cos\theta, \nonumber
\end{eqnarray}

\noindent for the region $r\leq R$. Here $P(R, r, \phi-\theta)$ 
is the Poisson kernel

\begin{eqnarray}
P(R, r, \phi-\theta)
=\frac{R^{2}-r^{2}}{R^{2}+r^{2}-2rR\cos(\phi-\theta)}.
\end{eqnarray}

\noindent  We interchange $R$ and $r$ in (\ref{Poissonintegral}) 
for the region $R<r$. One can easily prove these, for example following
Ref. \onlinecite{AJ}. For each term in (\ref{BiotSavartLundquist}), we 
shall first carry out integration over $z$ and decompose the resulting 
integral into two pieces corresponding to regions $r\leq R$ and $R<r$. 
Then we shall carry out integrations over $\phi$ using 
(\ref{Poissonintegral}). We shall end up with integrals over 
$r$ ($x=\nu r$) combining these again. The first integral

\begin{eqnarray}
& & I_{1}=-\frac{1}{4\pi} \int
\frac{zF_{\phi}(r)\vect{e}_{r}(\phi)}
{[R^{2}+r^{2}-2rR\cos(\phi-\theta)+z^{2}]^{3/2}} 
rdr d\phi dz, \\
& & \hspace*{4mm} =0 \nonumber
\end{eqnarray}

\noindent immediately vanishes upon integration over $z$.
The second term yields

\begin{eqnarray}
& & I_{2}=\frac{1}{4\pi} \int
\frac{RF_{z}(r)\vect{e}_{\theta}}
{[R^{2}+r^{2}-2rR\cos(\phi-\theta)+z^{2}]^{3/2}} 
rdr d\phi dz \\
& & \hspace*{4mm}
=\frac{1}{\nu} F_{0} \left[ X\int_{x=0}^{X} \frac{J_{0}(x)}{X^{2}-x^{2}}xdx
+X\int_{x=X}^{\infty} \frac{J_{0}(x)}{x^{2}-X^{2}}xdx \right]
\vect{e}_{\theta}.
\nonumber
\end{eqnarray}

\noindent The third integral is

\begin{eqnarray}
& & I_{3}=-\frac{1}{4\pi} \int
\frac{rF_{z}(r)\vect{e}_{\phi}}
{[R^{2}+r^{2}-2rR\cos(\phi-\theta)+z^{2}]^{3/2}} 
rdr d\phi dz \\
& & \hspace*{4mm}
=\frac{1}{\nu} F_{0} \frac{1}{X}
\int_{x=0}^{X} J_{0}(x) xdx
\vect{e}_{\theta}-I_{2}.
\nonumber
\end{eqnarray}

\noindent The fourth integral yields

\begin{eqnarray}
& & I_{4}=-\frac{1}{4\pi} \int
\frac{RF_{\phi}(r)\cos(\phi-\theta)\vect{e}_{z}}
{[R^{2}+r^{2}-2rR\cos(\phi-\theta)+z^{2}]^{3/2}} 
rdr d\phi dz \\
& & \hspace*{4mm}
=-\frac{1}{\nu} F_{0} \left[ \int_{x=0}^{X}\frac{J_{1}(x)}{X^{2}-x^{2}}x^{2}dx
+X^{2}\int_{x=X}^{\infty} \frac{J_{1}(x)}{x^{2}-X^{2}}dx \right]
\vect{e}_{z}.
\nonumber
\end{eqnarray}

\noindent The fifth integral is

\begin{eqnarray}
& & I_{5}=\frac{1}{4\pi} \int
\frac{rF_{\phi}(r)\vect{e}_{z}}
{[R^{2}+r^{2}-2rR\cos(\phi-\theta)+z^{2}]^{3/2}} 
rdr d\phi dz \\
& & \hspace*{4mm}
=\frac{1}{\nu} F_{0} \int_{x=X}^{\infty} J_{1}(x) dx 
\vect{e}_{z} -I_{4}.
\nonumber
\end{eqnarray}

  Thus we find  

\begin{eqnarray}
& & BS[\vect{F}_{L}](\vect{x})
=\frac{1}{\nu} F_{0} \left[ \frac{1}{X} \int_{x=0}^{X} J_{0}(x) xdx
\vect{e}_{\theta}+\int_{x=X}^{\infty} J_{1}(x) dx 
\vect{e}_{z} \right] \nonumber \\
& & \hspace*{19mm} =\frac{1}{\nu} 
\vect{F}_{L}(\vect{x}),
\end{eqnarray}

\noindent adding up these terms. Here, we use 

\begin{eqnarray}
\frac{1}{X} \int_{x=0}^{X} J_{0}(x)xdx=J_{1}(X),
\hspace*{8mm}
\int_{x=X}^{\infty} J_{1}(x)dx=J_{0}(X).
\end{eqnarray}

\end{document}